\documentstyle[11pt]{article}
\input epsf
\makeatletter
\renewcommand\theequation{\thesection.\arabic{equation}}
\@addtoreset{equation}{section}
\makeatother

\def\fun#1#2{\lower3.6pt\vbox{\baselineskip0pt\lineskip.9pt
  \ialign{$\mathsurround=0pt#1\hfil##\hfil$\crcr#2\crcr\sim\crcr}}}

\def\be{\begin{equation}}
\def\ee{\end{equation}}

\begin{document}
\baselineskip 20pt
\thispagestyle{empty}
\rightline{CU-TP-755}
\rightline{hep-th/9605229}
\vskip 2cm
\centerline{\Large\bf Massive and Massless Monopoles}
\centerline{\Large\bf with Nonabelian Magnetic Charges}
\vskip 1.5cm
\centerline{\large\it
Kimyeong Lee,\footnote{electronic mail: klee@phys.columbia.edu}
Erick J. Weinberg,\footnote{electronic mail: ejw@phys.columbia.edu} and 
Piljin Yi\footnote{electronic mail: piljin@phys.columbia.edu}}
\vskip 1mm
\centerline{Physics Department, Columbia University, New York, NY, 10027}
\vskip 1.5cm
\centerline{\bf ABSTRACT}
\vskip 5mm
\begin{quote}
{\baselineskip 16pt
We use the multimonopole moduli space as a tool for studying the properties of
BPS monopoles carrying nonabelian magnetic charges.  For configurations whose
total magnetic charge is purely abelian, the moduli space for nonabelian
breaking can be obtained as a smooth limit of that for a purely abelian
breaking.  As the asymptotic Higgs field is varied toward one of the special
values for which the unbroken symmetry is enlarged to a nonabelian group, some
of the fundamental monopoles of unit topological charge remain massive but
acquire nonabelian magnetic charges.  The BPS mass formula indicates that
others should become massless in this limit.  We find that these do not
correspond to distinct solitons but instead manifest themselves as
``nonabelian clouds'' surrounding the massive monopoles.  The moduli space
coordinates describing the position and $U(1)$ phase of these massless
monopoles are transformed into an equal number of nonabelian global gauge
orientation and gauge-invariant structure parameters characterizing the
nonabelian cloud.  We illustrate this explicitly in a class of $Sp(2N)$
examples for which the full family of monopole solutions is known.  We show in
detail how the unbroken symmetries of the theory are manifested as isometries
of the moduli space metric.  We discuss the connection of these results to the
Montonen-Olive duality conjecture, arguing in particular that the massless
monopoles should be understood as the duals to the massless gauge bosons that
appear as the mediators of the nonabelian forces in the perturbative
sector. 

}
\end{quote}

\newpage
\setcounter{footnote}{0}
\section{Introduction}

Magnetic monopoles have been the object of intense interest ever since
it was shown that they can arise as classical solutions in spontaneously 
broken gauge theories \cite{hooft}.  This interest is due in part to
their role as predicted, although as yet undiscovered, particles that
occur in all grand unified theories.  Beyond their specific
phenomenological implications, however, monopoles 
are of interest as examples of classical solitons.  Like all solitons,
they give rise after quantization to a type of particle that can be seen
as complementary to those that arise as quanta of the elementary
fields.  The complementary nature of solitons and elementary quanta is
particularly striking in theories with unbroken $U(1)$ gauge symmetry,
since Maxwell's equations are invariant under a duality that
interchanges magnetic and electric charges.  This idea is made more
concrete in the conjecture of Montonen and Olive \cite{dual} that in certain
theories there might be an exact electric-magnetic duality that
exchanges solitons and 
elementary quanta, and weak and strong coupling.

In this paper we will be concerned with monopoles whose magnetic
charge has a nonabelian component; i.e., those whose long-range
magnetic field transforms nontrivially under an unbroken nonabelian
subgroup of the gauge symmetry of the theory.  Just as elementary
quanta carrying nonabelian electric-type charges display a much richer
range of phenomena than those with purely abelian charges, there are
some curious new properties that arise with nonabelian magnetic
charges.   Some of these are associated with the long-distance behavior
of these monopoles.  Attempts to apply a time-dependent global nonabelian
gauge rotation to obtain a dyonic object carrying both electric and
magnetic nonabelian charges are frustrated by the nonnormalizability
of certain zero modes \cite{abouel} and, at a deeper level, by the inability 
to define global nonabelian charge in the presence of a monopole
\cite{manohar,nelson}.  Also, Brandt and Neri and Coleman \cite{brandt}
have shown that, regardless of the physics that governs the structure
of their core, monopoles carrying more than a minimal nonabelian
magnetic charge are unstable against decay into minimally charged
objects.  (This result does not apply in the BPS limit.)  
There are other new phenomena suggested by the possibility of
electric-magnetic duality.  In particular, one would expect the
massless electrically charged gauge bosons to have magnetic
counterparts.  Although duality would predict that these should be
massless, it is not obvious how to obtain a zero energy soliton; one
of our goals will be to gain more insight into the properties of these
objects.  

We work with an adjoint representation Higgs field $\Phi$ in the
Bogomol'nyi-Prasad-Sommerfield (BPS) limit \cite{bps} in which the
scalar field potential is ignored and a nonzero Higgs expectation
value is imposed as a boundary condition at spatial infinity.  In this
limit static monopole solutions obey the first order
equations\footnote{Our conventions are such that $F_{ij} = \partial_i
A_j -\partial_j A_i +i e[A_i,A_j] = \epsilon_{ijk} B_k$ and $D_i \Phi =
\partial_i \Phi + ie [A_i ,\Phi]$.}  
\begin{equation}
       B_i =  D_i \Phi\, .
\label{bogomolny}
\end{equation} 
Because the Higgs field is massless in the BPS limit, it mediates a
long-range force.  For static monopoles, this force 
exactly balances their magnetic force.

We also use the moduli space approximation 
\cite{geodesic}, in which the dynamics of the many
degrees of freedom of the soliton solution is effectively reduced to
that of a small number of collective coordinates $z_i$.  For BPS
monopoles, the absence of static interactions implies that the collective
coordinate Lagrangian consists only of a
kinetic energy term, which can be written in the form 
\begin{equation}
     L = {1 \over 2} g_{ij}(z)\, \dot z_i \dot z_j 
\end{equation} 
where $g_{ij}$ may be interpreted as a metric on the moduli space
spanned by the $z_i$.  If the monopole solutions are known for arbitrary
values of the collective coordinates, then the moduli space metric can be
obtained in a straightforward manner   from the zero modes about these
solutions.  Even if the general solution is not known, as is usually the
case, it is sometimes possible to determine the moduli space metric. 
This was first done by Atiyah and Hitchin \cite{atiyah}, who found the  
two-monopole moduli space metric for the case of $SU(2)$ broken to $U(1)$.   
Recently, the metric for two monopoles in a theory with an arbitrary 
group broken to a purely abelian subgroup was found [10-12].
Finally, in Ref.~\cite{klee2} we proposed a family of metrics 
for 
the moduli spaces of a somewhat larger 
class of multimonopole solutions in higher rank gauge
groups.\footnote{In a recent paper, Murray \cite{murray} has shown that these  
metrics coincide with those on the space of Nahm data 
\cite{nahm} for the unitary gauge groups. 
More recently, 
Chalmers \cite{chalmers} has given a proof that they
are the only smooth hyperk\"ahler metrics that possess the right
symmetry properties as well as the correct asymptotic behavior, and thus
are in fact the exact moduli space metrics.}

These last results are the starting point for our present
investigation.  We begin in Sec.~2 by reviewing some of the properties
of BPS monopoles.  An adjoint Higgs field can break a rank $r$ gauge
group $G$ to either $U(1)^r$ or to $K \times U(1)^{r-k}$, where $K$ is
a semisimple group of rank $k<r$. The former case, which we will refer
to as maximal symmetry breaking (MSB), occurs for generic values of
$\Phi$. There are $r$ topologically conserved charges, one for each $U(1)$
factor.  Associated with these are $r$ fundamental monopoles, each
carrying a single unit of one of these topological charges; all other
BPS solutions can be understood as multimonopole solutions containing
appropriate numbers of the various fundamental monopoles.  

The latter case, with a nonabelian unbroken symmetry (NUS), occurs for
special values of $\Phi$.  For these values some of the fundamental
monopoles of the MSB case survive as massive solitons but acquire
nonabelian magnetic charge in the sense that their long-range magnetic
field has nonvanishing components in $K$.  Taken at face value, the
BPS mass formulas indicate that certain other fundamental MSB
monopoles (also with nonabelian magnetic charge) become massless in
the NUS limit; these are just the duals to the massless gauge bosons
that were mentioned above.  Their interpretation is complicated by the
fact that as the massless limit is approached the core radii of the
corresponding classical monopole solutions tend to infinity while at
the same time the fields all tend toward their vacuum values.

In this paper we investigate the properties of these nonabelian
monopoles by following the behavior of MSB solutions as the asymptotic
Higgs field is varied toward the NUS value.  For configurations whose
total magnetic charge acquires a nonabelian component when one passes
from the MSB to the NUS case, one encounters various pathologies, of
which the behavior of the massless monopoles described above is just
one example.  To avoid these difficulties, we use the approach of
Refs.~\cite{nelson,hollo}  and consider only combinations of
monopoles whose nonabelian charges cancel.  As we shall see, for such
``magnetically color-neutral'' combinations the approach to the NUS
case is quite smooth.

Each such combination of NUS magnetic charges is the
limit of a unique set of MSB magnetic charges.  Index theory methods  
reveal that the moduli spaces for the two cases have the same
dimension.  It therefore seems quite plausible that the moduli space
metric for the NUS case should be simply the appropriate
limit of the 
metric for the corresponding set of MSB charges.  

In Sec.~3 we test this explicitly for an example with gauge group
$SO(5)$, with MSB and NUS symmetry breakings to $U(1)\times U(1)$ and 
$SU(2) \times U(1)$, respectively.  In the former case there are two
fundamental monopole solutions.  Because the sum of the magnetic charges
of these two remains purely abelian as one passes to the NUS case, the 
solutions containing two distinct fundamental monopoles are precisely
the sort of color-neutral combinations that we want.  For the MSB case,
the metric for the corresponding eight-dimensional two-monopole moduli
space is known from the results of Ref.~\cite{klee}.  For the NUS case, 
the full
eight-parameter family of solutions was found some time ago \cite{so5}.  We use
these to calculate the NUS moduli space metric directly and verify that
it is indeed the expected limit of the MSB metric.

Despite this smooth behavior of the metric, the interpretation of the
coordinates on the moduli space is quite different for the cases of
abelian and nonabelian symmetry breaking.  In the MSB case the
generic solution has a natural interpretation in terms of two widely
separated monopoles, each of which is specified by the three spatial
coordinates of its center and a single $U(1)$ phase angle.  As the NUS
limit is approached, one of the fundamental monopole solutions retains
its nonzero mass and finite core radius.  The mass of the other
fundamental monopole approaches zero while, as noted above, its
radius, {\it in the absence of any other monopoles,} tends to
infinity.  However, the behavior of this massless monopole is modified
dramatically by the presence of a massive monopole.
 
This can be seen by considering an MSB solution containing two such monopoles
separated by a distance $r_0$ that is much larger than either of their
core radii.  As the NUS limit is approached, the core of the would-be
massless monopole expands until its radius becomes comparable to
$r_0$.  It then begins to lose its identity as an isolated soliton and
instead is manifested as a ``nonabelian cloud'' of radius $\sim r_0$
surrounding 
the massive monopole.  Within this cloud there 
is a Coulomb magnetic field corresponding to a combination of abelian
and nonabelian magnetic charge, but the 
nonabelian component disappears for $r \gg r_0$.  In the NUS
nonabelian limit, one of the position coordinates of the massless
monopole is transformed into a parameter specifying the radius of
the nonabelian cloud, while its other two position coordinates combine
with its $U(1)$ phase angle to specify the global $SU(2)$ orientation
of the solution.

In the last part of Sec.~3, we consider the semiclassical quantization
of the moduli space coordinates describing this nonabelian cloud.  We
find that there is a tower of states carrying both spin and electric-type
$SU(2)$ charge (``isospin''), with the magnitudes of
the isospin and spin being equal.

In Sec.~4 we consider some more complex cases. The first of these involves 
a color-neutral combination of $(N+1)$ monopoles in a theory with $Sp(2N+2)$ 
broken to $Sp(2N) \times U(1)$. (For $N=1$ this reduces to the $SO(5)=Sp(4)$ 
example of Sec.~3.) The $N+1$ monopoles become distinct fundamental 
monopoles upon maximal symmetry breaking, so the MSB moduli space metric 
given in Ref.~\cite{klee2} is applicable to this case.  
With $Sp(2N)\times U(1)$ as the unbroken group, $N$ of these monopoles become 
massless and coalesce in a cloud about the single massive monopole.  In fact, 
the full family of solutions for this case can be obtained from embedding of 
the $SO(5)$ solutions of section III.  As with the $SO(5)$ case, one can 
verify that the moduli space metric obtained from such exact monopole 
solutions is identical to the NUS limit of the MSB moduli space metric. 
This example also illustrates very nicely how monopole 
coordinates are transformed into parameters describing the structure
and gauge orientation of the cloud. As we will 
show, what used to be the relative position and $U(1)$ coordinates of the 
$N+1$ monopoles can be assembled into $2N$ complex (or $N$ quaternionic) 
variables on which the unbroken group $Sp(2N)$ acts
triholomorphically, 
defining a set of Killing vector fields of the algebra 
of $Sp(2N)$. These leave invariant a single combination of the monopole 
coordinates that becomes the radius of the nonabelian cloud.

The next step is to examine solutions with two massive monopoles in
the NUS limit.  For either $Sp(2N+2)$ or $SU(N+2)$ broken to $SU(N)
\times U(1)^2$ there are magnetically color-neutral configurations
with $(N-1)$ massless and two massive monopoles, each of which
individually carries a nonzero nonabelian magnetic charge. Again they
belong to the class of multimonopoles for which we have a 
MSB moduli space metric. We are unable to compare its NUS limit 
to the exact metric in this case, because the complete family of such
multimonopole solutions is unknown. Instead, we examine its symmetry
properties in the NUS limit, which must include a $U(N)$
triholomorphic isometry coming from the unbroken gauge group, and find
the right set of Killing vectors.  As in the previous case, we can
construct a single invariant from the massless monopoles coordinates
that fixes the size of the nonabelian cloud surrounding the two
massive monopoles.

We cannot carry out the analysis at this level for other cases, since
we know neither the general solutions nor the moduli space metric.
However, as we describe in Sec.~5, it is still possible to make some
progress in understanding nonabelian monopoles in other groups.  From
the root structure of the group, we can determine the transformation
properties of the massive fundamental monopoles under the unbroken
gauge group and see how they can be combined to yield configurations
with no net nonabelian magnetic charge.  Each such combination
requires a fixed number of massless monopoles, whose coordinates
combine to give the various global gauge and cloud structure
parameters.  Using group theory arguments, we can in most cases
determine (and in the remaining ones bound) the number of structure
parameters.  In general there are more than one, suggesting that the
nonabelian cloud can have considerable structure.

One of motivations for this work was the possibility of an exact
electric-magnetic duality.  In particular, it has been conjectured
that in $N=4$ supersymmetric Yang-Mills theories there is a
correspondence between electrically and magnetically charged states.
While some of the magnetic states required by this duality are based
straightforwardly on the fundamental monopole solutions, others must
be obtained as threshold bound states; the latter can be related to
normalizable harmonic forms on the moduli space.  In Sec.~6 we
note some of the implications of our results for this conjectured
duality and discuss some of the issues related to the threshold bound
states.

Finally, in Sec.~7 we summarize our results and make some concluding
remarks.  Some detailed calculations relating to the isometries of the
moduli spaces studied in Sec.~4.3 are contained in the Appendix.

\section{Review of BPS monopoles}

We begin by recalling the main features of the BPS monopoles in an
$SU(2)$ gauge theory \cite{bps}.  We fix the normalization of the magnetic
charge in the unbroken $U(1)$ by writing the asymptotic magnetic field
as 
\begin{equation}
    B_i^a =  {g{\hat r}_i \over 4\pi r^2} {\Phi^a \over |\Phi|}.
\end{equation}
Topological arguments then show that $g$ must be quantized in integer
multiples of $4\pi/e$.  The monopole solution carrying one unit of
magnetic charge may be written as
\begin{eqnarray}
   \Phi^a &=&  {\hat r}^a H(r),   \nonumber \\
   A_i^a  &=&  \epsilon_{aim}{\hat r}^m A(r),
\end{eqnarray}
where $v$ is the asymptotic magnitude of the Higgs field and 
\begin{eqnarray}
   A(r) &=& {v \over \sinh evr} - {1\over er}, \nonumber \\
   H(r) &=& v \coth evr  - {1\over er}.
\label{AHdef}
\end{eqnarray}
The solutions carrying $n > 1$ units of magnetic charge can all be
understood as multimonopole solutions.  The dimension of the moduli
space for a given $n$ can be determined by
studying the zero modes about an arbitrary solution; i.e, the
perturbations that preserve 
Eq.~(\ref{bogomolny}) to first order.  By requiring that these
perturbations satisfy the background gauge condition
\begin{equation}
    0 = D_i \delta A_i + ie[\Phi, \delta \Phi]  \equiv D_\mu \delta A_\mu,
\end{equation}
we ensure that the zero mode is orthogonal to all modes obtained by
gauge transformation of the original solution with gauge functions
that vanish at spatial infinity\footnote{In the second equality we
have adopted a notation in which $\Phi$ is treated as the fourth
component $A_4$ of a vector potential $A_\mu$, with $\partial_4$
acting on any quantity being identically zero.  We will always use
Greek indices to indicate that this four-dimensional notation is being
used; Roman indices should always be understood to run from 1 to 3.}.
This leaves only a single normalizable gauge mode, corresponding to
the single generator of the unbroken $U(1)$.  Index theory methods
show \cite{erick} that there are $4n$ linearly independent normalizable 
zero modes; when the monopoles are separated far away from each other,
the corresponding coordinates on the moduli space
having natural interpretations as the positions and $U(1)$ phase
angles of $n$ unit monopoles. 

    Now consider an arbitrary gauge group $G$ of rank $r$.  Its
generators can be chosen to be $k$ commuting operators $H_i$,
normalized by ${\rm tr}\,H^iH^j=\delta^{ij}$, that span the Cartan
subalgebra, together with ladder operators, associated with the roots
${\mbox{\boldmath $\alpha$}}$, that obey
\begin{equation}
     [{\bf H}, E_{\mbox{\boldmath $\alpha$}}] = {\mbox{\boldmath $\alpha$}} 
E_{\mbox{\boldmath $\alpha$}},  \qquad \qquad  
    [E_{\mbox{\boldmath $\alpha$}}, E_{-{\mbox{\boldmath $\alpha$}}}] = 
{\mbox{\boldmath $\alpha$}} \cdot {\bf H} \, .
\end{equation}
One can choose a basis of $r$ simple roots with the property that all
other roots are linear combinations of these with integer
coefficients all of the same sign.  A particularly convenient basis
may be chosen as follows.  Let $\Phi_0$ be the asymptotic value of
$\Phi$ in some fixed direction.  We 
choose this to lie in the Cartan subalgebra and then define a vector
$\bf h$ by
\begin{equation}
           \Phi_0 = {\bf h \cdot H}.
\end{equation}
We then require that the simple roots all have nonnegative inner
products with $\bf h$. If the symmetry breaking is maximal,
there are no roots orthogonal to $\bf h$ and there is 
unique set of simple roots ${\mbox{\boldmath $\beta$}}_a$ obeying 
this condition.   If instead there are roots
orthogonal to $\bf h$, then the sublattice formed by such roots is the
root lattice for some semisimple group $K$ of rank $k<r$, and the
unbroken gauge group is $U(1)^{r-k}\times K$.  In this case we denote by
${\mbox{\boldmath $\gamma$}}_j$ the simple roots orthogonal to $\bf h$ and 
write the remainder as ${\mbox{\boldmath $\beta$}}_a$.  
Here the choice of simple roots is not unique, with the various
possibilities being related by elements of the Weyl group of $K$.

We can also require that, in the direction chosen to define $\Phi_0$,
the asymptotic magnetic field lie in the Cartan subalgebra and be
of the form 
\begin{equation}
    B_i = {{\hat r}_i \over 4\pi r^2} {\bf g\cdot H}.
\end{equation}
Topological arguments lead to the quantization condition \cite{topology}
\begin{equation}
    {\bf g} = {4\pi \over e} \left[\sum_{a} n_a 
{\mbox{\boldmath $\beta$}}_a^* 
         + \sum_{j} q_j {\mbox{\boldmath $\gamma$}}_j^* \right],
\label{gcoeff}
\end{equation} 
where  
\begin{equation}
    {\mbox{\boldmath $\alpha$}}^* = {{\mbox{\boldmath $\alpha$}} \over 
{\mbox{\boldmath $\alpha$}}^2}
\end{equation} 
is the dual of the root ${\mbox{\boldmath $\alpha$}}$ and the $n_a$ and 
$q_j$ are non-negative
integers.  The $n_a$ are the topologically conserved charges.  For a given
solution they are uniquely determined and gauge invariant, even though 
the corresponding ${\mbox{\boldmath $\beta$}}_a$ may not be.
The $q_j$ are neither gauge invariant nor conserved.

For maximal symmetry breaking there is a unique fundamental
monopole solution associated with each of the $r$ topological charges.
To obtain these, we first note that any root ${\mbox{\boldmath $\alpha$}}$ 
defines an $SU(2)$ subgroup generated by
\begin{eqnarray}
t^1({\mbox{\boldmath $\alpha$}}) &=& \frac{1 }{ \sqrt{2
{\mbox{\boldmath $\alpha$}}^2}} (E_{\mbox{\boldmath $\alpha$}} +
E_{-{\mbox{\boldmath $\alpha$}}})       ,           \nonumber \\
t^2({\mbox{\boldmath $\alpha$}}) &=& -\frac{i }{ \sqrt{2{
\mbox{\boldmath $\alpha$}}^2}} (E_{\mbox{\boldmath $\alpha$}} -
E_{-{\mbox{\boldmath $\alpha$}}})         ,         \nonumber \\
t^3({\mbox{\boldmath $\alpha$}}) &=&  {\mbox{\boldmath $\alpha$}}^* 
\cdot  {\bf H}  .
\label{tripletdef}
\end{eqnarray}
If $A^s_i({\bf r}; v)$ and $\Phi^s({\bf r}; v)$ give the $SU(2)$
solution corresponding to a Higgs expectation value $v$, then the
fundamental monopole corresponding to the root ${\mbox{\boldmath $\beta$}}_a$ 
is given by \cite{erick}
\begin{eqnarray}
   A_i({\bf r})  &=&  \sum_{s=1}^3 A_i^s({\bf r}; {\bf h}\cdot 
{\mbox{\boldmath $\beta$}}_a) t^s({\mbox{\boldmath $\beta$}}_a) , \nonumber \\
   \Phi({\bf r})  &=&  \sum_{s=1}^3 \Phi^s({\bf r}; {\bf h}\cdot 
{\mbox{\boldmath $\beta$}}_a) t^s({\mbox{\boldmath $\beta$}}_a)   
   + ({\bf h} - {\bf h}\cdot {\mbox{\boldmath $\beta$}}_a^* 
{\mbox{\boldmath $\beta$}})\cdot {\bf H} .
\label{embedding}
\end{eqnarray}
It carries topological charges
\begin{equation}
     n_b = \delta_{ab}   ,
\end{equation}
and has mass
\begin{equation}
     m_a ={4\pi \over e} {\bf h}\cdot {\mbox{\boldmath $\beta$}}_a^* .
\end{equation}

All other BPS solutions can be understood as multimonopole solutions 
containing $N= \sum n_a$ fundamental monopoles.  These include both
solutions, containing many widely separated fundamental monopoles, that are
obviously composite and spherically symmetric solutions whose compositeness
is revealed only by analysis of their zero modes.  The latter
solutions are obtained
by replacing ${\mbox{\boldmath $\beta$}}_a$ in Eq.~(\ref{embedding}) by any 
composite root ${\mbox{\boldmath $\alpha$}}$; their 
topological charges are equal to the coefficients in the expansion 
\begin{equation}
    {\mbox{\boldmath $\alpha$}}^* = \sum_a  n_a {\mbox{\boldmath $\beta$}}_a^*.
\end{equation}

The moduli space for these multimonopole solutions has $4N$
dimensions, corresponding to three position variables and a single
$U(1)$ phase for each of the component fundamental monopoles.  The 
full moduli space and its metric are known for $N=2$.  
For $N>2$ the metric for the case where all the component fundamental
monopoles are all distinct was given in Ref.~\cite{klee2}; for all
other cases the explicit form of the metric is known only for the
region of moduli space corresponding to widely separated fundamental
monopoles.

Matters are somewhat more complicated when the unbroken gauge group is
nonabelian \cite{erick2}.  If ${\mbox{\boldmath $\beta$}}_a\cdot \bf
H$ commutes with the generators of $K$ (i.e., if ${\mbox{\boldmath
$\beta$}}_a$ is not linked in the Dynkin diagram to one of the
${\mbox{\boldmath $\gamma$}}_j$), the construction described above
yields a unique fundamental monopole carrying a single unit of
topological charge.  The identification of the fundamental solutions
for the remaining ${\mbox{\boldmath $\beta$}}_a$ is less
straightforward.  The Weyl group of $K$ takes each of these
${\mbox{\boldmath $\beta$}}_a$ to one or more other roots, any of
which could have been chosen as a simple root instead of
${\mbox{\boldmath $\beta$}}_a$.  Using any of these in the embedding
construction leads to a solution, carrying a single unit of
topological charge, that is simply a global gauge rotation of the
original solution.  In addition, it is sometimes possible to have a root
${\mbox{\boldmath $\alpha$}}$ that is not related to ${\mbox{\boldmath
$\beta$}}_a$ by a Weyl reflection but that nevertheless gives an
expansion
\begin{equation}
     {\mbox{\boldmath $\alpha$}}^* = {\mbox{\boldmath $\beta$}}^*_a + 
\sum_j q_j {\mbox{\boldmath $\gamma$}}_j^* .
\end{equation}
Insertion of such a root into Eq.~(\ref{embedding}) yields a solution that is
gauge-inequivalent to the solution based on ${\mbox{\boldmath $\beta$}}_a$, 
yet still carries unit topological charge.\footnote{Such solutions were
referred to as degenerate fundamental monopoles in Ref.~\cite{erick2}}  
As we will see illustrated in the next section, there is a continuous family of
gauge-inequivalent solutions with unit topological charge that
interpolate between the ${\mbox{\boldmath $\alpha$}}$- and 
${\mbox{\boldmath $\beta$}}_a$-embedding solutions.    

If the long-range magnetic field has a nonabelian component (i.e., if
${\bf g}\cdot {\mbox{\boldmath $\gamma$}}_j)\ne 0$, the index theory methods 
used to count zero modes in Refs.~\cite{erick} and \cite{erick2} fail for 
technical reasons related to the slow falloff of the nonabelian field at 
large distance.  These difficulties do not arise if ${\bf g}\cdot
{\mbox{\boldmath $\gamma$}}_j =0$, in which case the
number of normalizable zero modes is
\begin{equation}
    p = 4 \left[ \sum_a n_a + \sum_j q_j \right] .
\end{equation}
(It is possible to write $p$ in the form $\sum
c_a n_a$, but this is somewhat misleading because, as we will see, there 
are some zero modes that cannot be associated with any single
fundamental monopole.)

\section{An $SO(5)$ example}
\smallskip
\subsection{Monopoles in $SO(5)$ Gauge Theory}
\smallskip

Many of the issues we want to address are illustrated in a
particularly simple fashion if the gauge group $G$ is $SO(5)$, whose
root lattice is shown in Fig.~1. If $\bf h$ is oriented as shown in Fig.~1a, 
there is maximal symmetry breaking, to the subgroup $U(1) \times U(1)$, 
while if ${\bf h}$ is as in Fig.~1b, the unbroken gauge group is $SU(2) 
\times U(1)$ with the $SU(2)$ defined by the long root 
$\mbox{\boldmath $\gamma$}$.\footnote{There 
is an inequivalent breaking to $SU(2) \times U(1)$ where the unbroken 
$SU(2)$ is the subgroup defined by a short root; this case is not 
of interest to us here.}  
In this section we will examine the behavior 
as $\bf h$ is rotated toward ${\mbox{\boldmath $\alpha$}}$ (i.e., as the
mass of the ${\mbox{\boldmath $\gamma$}}$ vector meson is taken to zero) and 
see to what extent the properties of the monopoles with $SU(2) \times U(1)$ 
symmetry breaking can be obtained as limits of the MSB case.

In the maximally broken case, with $\bf h$ oriented as in Fig.~1a, the simple
roots are the two labeled ${\mbox{\boldmath $\beta$}}$ and 
${\mbox{\boldmath $\gamma$}}$. The corresponding fundamental monopoles, 
with masses 
\begin{eqnarray}
  m_{\mbox{\boldmath $\beta$}}  &=& {4\pi \over e} {\bf h}\cdot 
{\mbox{\boldmath $\beta$}}^* , \nonumber \\ 
  m_{\mbox{\boldmath $\gamma$}}  &=& {4\pi \over e} {\bf h}\cdot 
{\mbox{\boldmath $\gamma$}}^* ,
\end{eqnarray}
are obtained by $SU(2)$ embeddings as in Eq.~(\ref{embedding}).
Their central cores have radii 
\begin{eqnarray}
  R_{\mbox{\boldmath $\beta$}}  &\sim& ( e{\bf h}\cdot 
{\mbox{\boldmath $\beta$}})^{-1}, \nonumber \\ 
  R_{\mbox{\boldmath $\gamma$}}  &\sim& (e {\bf h}\cdot 
{\mbox{\boldmath $\gamma$}})^{-1}
\label{radii}
\end{eqnarray}
that are set by the masses of the corresponding electrically charged
vector bosons.   

The $SU(2)$ embeddings defined by ${\mbox{\boldmath $\alpha$}}$ and 
${\mbox{\boldmath $\mu$}}$ give two other spherically symmetric solutions 
but, as discussed in Sec.~2, these are actually multimonopole solutions.
Because
\begin{eqnarray}
    {\mbox{\boldmath $\alpha$}}^* &=& {\mbox{\boldmath $\beta$}}^* + 
{\mbox{\boldmath $\gamma$}}^* ,  \nonumber \\ 
    {\mbox{\boldmath $\mu$}}^* &=& {\mbox{\boldmath $\beta$}}^* + 
2 {\mbox{\boldmath $\gamma$}}^* ,  
\label{superpositions}
\end{eqnarray}
the former is a two-monopole solution that maps to a single point of an
eight-dimensional moduli space, while the latter is a three-monopole
configuration, with the corresponding moduli space having twelve
dimensions.  Note that, even though these last two solutions are
composite, their cores are actually smaller than those of either of the
fundamental monopoles.  Essentially, this is because the vector boson mass
that sets the core size depends on the local, rather than the
asymptotic, value of the Higgs field.
\vskip 5mm
\begin{center}
\leavevmode
\epsfysize =3in\epsfbox{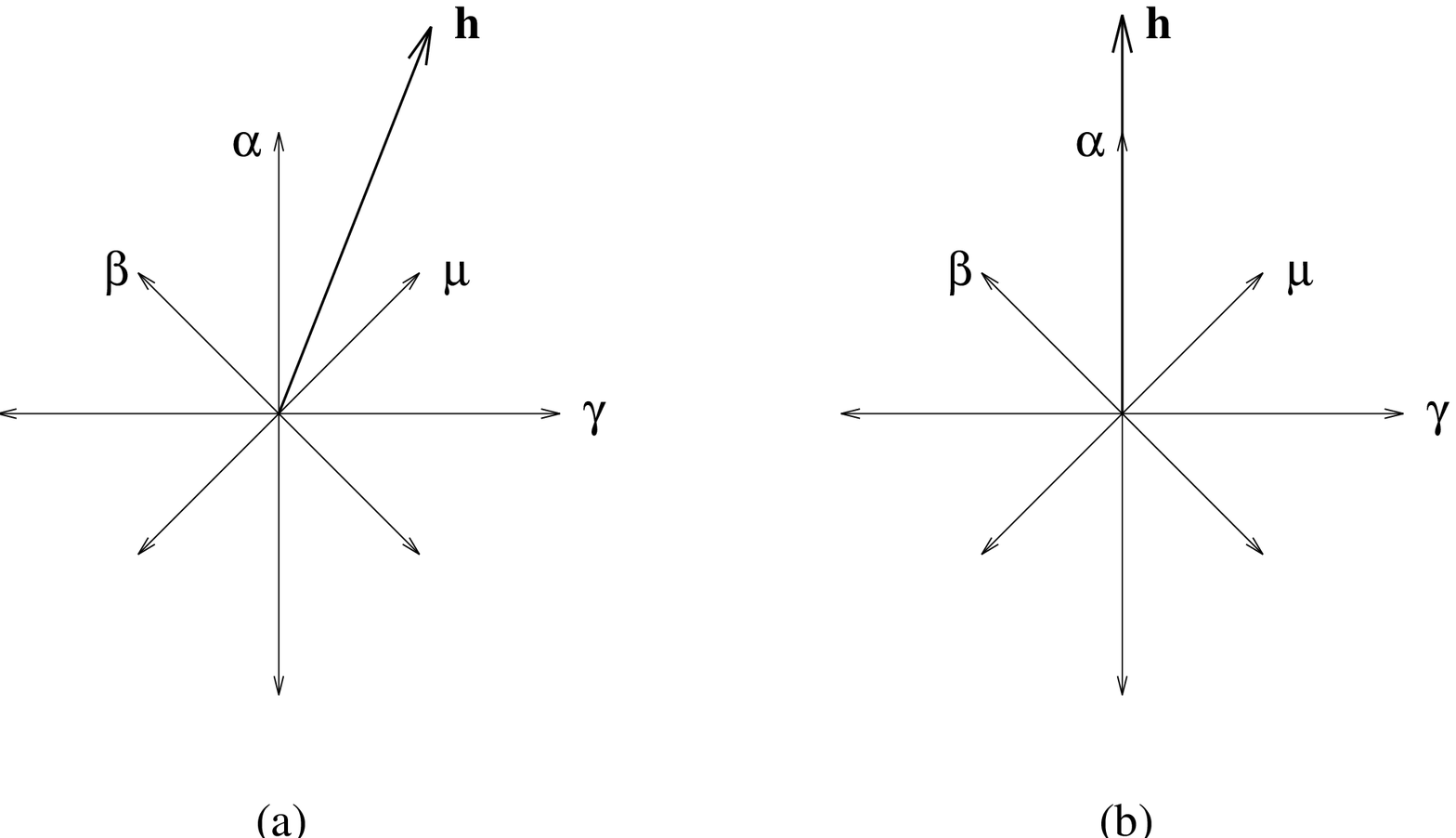}
\end{center}
\vskip 0mm
\begin{quote}
{\bf Figure 1:} {\small 
The unitary gauge Higgs expectation value $\bf h$
in the root space of $SO(5)$. The symmetry is maximally broken for (a) and
only partially for (b).}
\end{quote}
\vskip 1cm

With the nonabelian symmetry breaking that results when $\bf h$ is
orthogonal to ${\mbox{\boldmath $\gamma$}}$, as in Fig.~1b, 
${\mbox{\boldmath $\beta$}}$ and ${\mbox{\boldmath $\gamma$}}$ can
still be chosen as the simple roots.  However, there is no longer a
solution with $\bf g$ parallel to ${\mbox{\boldmath $\gamma$}}$.  
Furthermore, the solutions with $e{\bf g}/4\pi$ equal to 
${\mbox{\boldmath $\beta$}}^*$, ${\mbox{\boldmath $\beta$}}^* +
{\mbox{\boldmath $\gamma$}}^*$, and ${\mbox{\boldmath $\beta$}}^* + 2
{\mbox{\boldmath $\gamma$}}^*$ that corresponded to one, two,
and three monopoles in the MSB case are now all degenerate, with any
solution with $e{\bf g}/4\pi = {\mbox{\boldmath $\beta$}}^* +
2{\mbox{\boldmath $\gamma$}}^*$ being
gauge-equivalent to one with $e{\bf g}/4\pi = {\mbox{\boldmath $\beta$}}^*$.

The way in which this behavior emerges from the MSB case as $\bf h$ is
rotated toward ${\mbox{\boldmath $\alpha$}}$ can be rather subtle.  Consider, 
for example, the ${\mbox{\boldmath $\gamma$}}$ monopole solution.  This 
exists for all nonzero values of $m_{\mbox{\boldmath $\gamma$}}$, but not if 
$m_{\mbox{\boldmath $\gamma$}}=0$.  As $m_{\mbox{\boldmath $\gamma$}}$
decreases, the core of this monopole spreads out to increasingly large 
distances, while the magnitudes of the gauge fields at any fixed point 
in the core become ever smaller.  Thus, to an observer who measures fields 
only within a fixed region of space, the monopole becomes effectively 
undetectable when $m_{\mbox{\boldmath $\gamma$}}$ is sufficiently small.  
From a more global point of view,
on the other hand, the limit is not smooth.  Similarly, since the
moduli spaces for solutions with $e{\bf g}/4\pi={\mbox{\boldmath $\beta$}}^*$ 
and $e{\bf g}/4\pi={\mbox{\boldmath $\beta$}}^* +2
{\mbox{\boldmath $\gamma$}}^*$ have four and twelve dimensions,
respectively, they cannot have a common limit, even though an observer
confined to a finite volume would not be able to distinguish between
the case where $m_{\mbox{\boldmath $\gamma$}}$ was precisely zero and 
that where it very small, but nonzero.

Since these difficulties are associated, at least in part, with the
appearance of a nonabelian magnetic charge with its associated
Coulomb field, one
hope that the $m_{\mbox{\boldmath $\gamma$}}\rightarrow 0$ limit would 
be smoother for the solutions with $e{\bf g}/4\pi = 
{\mbox{\boldmath $\beta$}}^* + {\mbox{\boldmath $\gamma$}}^*$, whose
magnetic charge remains purely abelian.
Index theory methods can be applied to these
solutions for either maximal or non-maximal symmetry breaking, and in
both cases show that the moduli space is eight-dimensional.  It thus
seems quite plausible that the moduli space metric for the latter case
might be the $m_{\mbox{\boldmath $\gamma$}}\rightarrow 0$ limit of the moduli 
space for the former.  To test this conjecture, we will obtain the
moduli space metric for the NUS case directly from the explicitly
solutions that were found in Ref.~\cite{so5}, and then compare this
with the $m_{\mbox{\boldmath $\gamma$}}\rightarrow 0$ limit of the MSB
metric that was obtained in Refs.~\cite{klee2}.

\subsection{An Eight-Parameter Family of Solutions}

We begin by describing the solutions of Ref.~\cite{so5}.  We start
with some notation.  For any hermitian element $P$ of the
Lie algebra we define two real vectors ${\bf P}_{(1)}$ and 
${\bf P}_{(2)}$ and $2 \times 2$ matrix $P_{(3)}$ obeying $P_{(3)}^* = -
\tau_2 P_{(3)} \tau_2$ by 
\begin{equation}
   P = {\bf P}_{(1)} \cdot {\bf t}({\mbox{\boldmath $\alpha$}}) 
       + {\bf P}_{(2)} \cdot {\bf t}({\mbox{\boldmath $\gamma$}}) 
       + {\rm tr}\, P_{(3)} M ,
\end{equation}
where ${\bf t}({\mbox{\boldmath $\alpha$}})$ and ${\bf t}
({\mbox{\boldmath $\gamma$}})$ are defined as in
Eq.~(\ref{tripletdef}) and 
\begin{equation}
  M = \frac{i }{ \sqrt{{\mbox{\boldmath $\beta$}}^2}} 
       \left(\matrix{E_{\mbox{\boldmath $\beta$}}  & 
      -E_{-{\mbox{\boldmath $\mu$}}} \cr
       E_{\mbox{\boldmath $\mu$}}  & \,\,\,
       E_{-{\mbox{\boldmath $\beta$}}}} \right).
\label{matrixdef}
\end{equation}
Note that a $2\pi$ gauge rotation generated by
any of the $t^a({\mbox{\boldmath $\gamma$}})$ changes the sign of 
$P_{(3)}$ but leaves the other components of $P$ invariant. 
The commutation relations of the generators imply that the components
of $R = [P,Q]$ are
\begin{eqnarray}
  {\bf R}_{(1)} &=&  i{\bf P}_{(1)} \times  {\bf Q}_{(1)}
      -\, {\rm tr}\, P_{(3)}^\dagger {\mbox{\boldmath $\tau$}} Q_{(3)} ,
   \nonumber\\
  {\bf R}_{(2)} &=&  i{\bf P}_{(2)} \times  {\bf Q}_{(2)}
      -\, {\rm tr}\, P_{(3)} {\mbox{\boldmath $\tau$}} Q_{(3)}^\dagger ,
   \nonumber\\
   R_{(3)} &=& \frac{1}{2} \left[
    {\bf P}_{(1)} \cdot{\mbox{\boldmath $\tau$}} Q_{(3)} - Q_{(3)} 
   {\mbox{\boldmath $\tau$}} \cdot{\bf P}_{(2)} 
 -  {\bf Q}_{(1)} \cdot{\mbox{\boldmath $\tau$}} P_{(3)} + P_{(3)} 
    {\mbox{\boldmath $\tau$}} \cdot{\bf Q}_{(2)}     
   \right] .
\end{eqnarray}

The family of spherically symmetric solutions found in \cite{so5} can be
written as 
\begin{eqnarray}
   A^a_{i(1)} &=& \epsilon_{aim}{\hat r_m} A(r)  \qquad\qquad 
   \phi^a_{(1)} = {\hat r_a} H(r) \nonumber \\
   A^a_{i(2)} &=& \epsilon_{aim}{\hat r_m} G(r) \qquad\qquad
   \phi^a_{(2)} = {\hat r_a} G(r) \nonumber \\
   A_{i(3)} &=& \tau_i F(r) \qquad\qquad
   \phi_{(3)} = -i I F(r) .
\label{ansatz}
\end{eqnarray}
Here $A(r)$ and $H(r)$ are the $SU(2)$
monopole functions given in Eq.~(\ref{AHdef}), while the other two
coefficient functions obey
\begin{eqnarray}
   0 &=& G' + (eG +{2\over r}) G + 4e F^2 ,\label{Gde}   \\
   0 &=& F' + \frac{e}{2} (H-2A+G) F ,  \label{Fde}
\end{eqnarray}  
together with the boundary conditions $G(0)=F(\infty) =G(\infty) =0$. 
There is no constraint on $F(0)$, although the gauge freedom noted below
Eq.~(\ref{matrixdef}) can be used to make it positive.   These
equations have a one 
parameter family of solutions 
\begin{eqnarray}
   F(r) &=& { v \over \sqrt{8} \cosh (evr/2) }  L(r, a)^{1/2}, \nonumber \\
   G(r) &=& A(r) L(r, a), \label{cloud}
\end{eqnarray}
where
\begin{equation}
   L(r, a) = \left[ 1 +  (r/a) \coth(evr/2) \right]^{-1} 
\end{equation}
and the parameter\footnote{This is related to the 
parameter $b =F(0)$ used in Ref.~\cite{so5} by $ev a =
16b^2/(1-8b^2)$.} $a$ has the dimension of length and 
ranges from 0 to $\infty$.  In these formulas $v =
{\bf h}\cdot {\mbox{\boldmath $\alpha$}}$. 

When $a=0$ the monopole is invariant under the unbroken $SU(2)$, since
the doublet and triplet components of the fields, proportional to
$F(r)$ and $G(r)$, vanish identically.  If $a \ne 0$ these components
are nonvanishing and can be thought of as constituting a ``nonabelian
cloud'' about the monopole.  The effect of $a$ on the long-range tail
of $G(r)$ is particularly striking.  For $ 1/ev
\mathrel{\mathpalette\fun <} r \mathrel{\mathpalette\fun <} a$, this
falls as $1/r$, thus yielding the Coulomb magnetic field appropriate
to a nonabelian magnetic charge.  At larger distances, however, the
falloff increases to $1/r^2$, showing that the magnetic charge is
actually purely abelian.  Not surprisingly, the limit $a \rightarrow
\infty$ gives a solution that is a gauge transformation of the
${\mbox{\boldmath $\beta$}}$-embedding of the $SU(2)$ monopole, for
which $\bf g$ actually does have a nonabelian component.

With the MSB case in mind, one might think of these solutions as being
superpositions of a ${\mbox{\boldmath $\beta$}}$ monopole and a 
${\mbox{\boldmath $\gamma$}}$ monopole.  The fact that it has a finite
core radius, even though Eq.~(\ref{radii}) gives $R_{\mbox{\boldmath
$\gamma$}}=\infty$ in the NUS limit, can be seen as analogous to the
contraction of the cores in the analogous MSB superposition that was
noted below Eq.~(\ref{superpositions}).

     This one-parameter family of solutions can be extended to an
eight-parameter family by the action of the symmetries of the theory.
Three of the additional parameters correspond to spatial translations
of the solution, while the remaining four are obtained by applying
global $SU(2) \times U(1)$ transformations generated by ${\bf
t}({\mbox{\boldmath $\gamma$}})$ and $t^3({\mbox{\boldmath $\beta$}})$.

\subsection{Zero Modes}

The moduli space metric can be obtained directly from the zero modes
about these solutions, provided that these modes satisfy the
background gauge condition $D_\mu \delta A_\mu = 0$.  In order to
satisfy this condition, it may be necessary to add an infinitesimal
gauge transformation to the zero
modes obtained by varying the parameters in the solution, so that the zero
mode corresponding to a collective coordinate $z$
will in general take the form
\begin{equation}
    \delta_z A_\mu = \partial_z A_\mu  + D_\mu \epsilon_z .
\end{equation}
Once these zero modes have been found, the moduli space metric is
given by 
\begin{equation}
    g_{ij} = \int d^3x \,{\rm tr}\, 
         \left(\delta_i A_\mu \,\,\delta_j A_\mu \right) .
\label{modetometric}
\end{equation}

    The determination of the zero modes is simplified considerably by the
fact that one zero mode can be used to generate three others.  If we
define 
\begin{equation}
     \psi(x) = I\delta \phi(x) + i \sigma_j \delta A_j(x) ,
\end{equation}
then the three self-duality equations plus the background gauge condition
for $\delta A_\mu$ are equivalent to the Dirac equation
\begin{equation}
      \sigma_\mu D_\mu \psi =  0 ,
\end{equation} 
where $\sigma_4 \equiv i$ \cite{lowell}.  Right
multiplication of a solution $\psi$ 
by any unitary $2 \times 2$ matrix yields another solution
$\psi'$, which can be transformed back to give a new zero mode $\delta'
A_\mu$.  In particular, if we have a zero mode $\delta A_\mu$, then
multiplication of the corresponding $\psi$ on the right by  
$i \hat{\bf n} \cdot {\mbox{\boldmath $\sigma$}}$ (where $\hat {\bf n}$ is a 
unit three-vector) yields a new Dirac solution that can be decomposed to give 
\begin{eqnarray}
    \delta' \phi &=&   -\hat n_i \delta  A_i  ,   \nonumber \\
    \delta' A_i &=& \hat n_i \delta  \phi  
      + \epsilon_{ijk}  \hat n_j \delta  A_k .
\label{newmode}
\end{eqnarray}
By making three orthogonal choices for $\hat {\bf n}$, we can obtain three
zero modes that are orthogonal to each other and to the original mode; the
four modes clearly have the same norm.  
 
    We consider first the mode corresponding to an infinitesimal change
in the  parameter $a$.  Because $a$ enters only through the
function $L$, 
\begin{equation}
   \partial_a A_{\mu (1)} = 0, \qquad  
   \partial_a A_{\mu (2)} =  {\partial_a L\over L}  A_{\mu (2)}, \qquad 
   \partial_a A_{\mu (3)} =  {\partial_a L\over 2L}  A_{\mu (3)}.
\label{amode}
\end{equation}
To see whether this is already in background gauge, we must calculate 
\begin{equation}
    D_\mu \partial_a A_\mu = \partial_\mu \delta A_\mu 
        + ie[A_\mu, \delta A_\mu ] .
\end{equation}
It is trivial to verify the vanishing of the singlet and triplet
components of this quantity.  The remaining component is 
\begin{equation}
    D_\mu \partial_a A_{\mu(3)} = \partial_j \partial_a A_{j(3)} 
      +{i e\partial_a L \over 4L} \left[ 
   {\bf A}_{\mu(1)} \cdot {\mbox{\boldmath $\tau$}} A_{\mu(3)} + 
    A_{\mu(3)} {\mbox{\boldmath $\tau$}} \cdot {\bf A}_{\mu(2)} \right] .
\end{equation} 
In the first term on the right we can interchange the spatial 
differentiation and the variation of $a$.  To evaluate the 
second term we make use of the fact that $\partial_a L/ L = 2\partial_a
F/F $.  Using Eqs.~(\ref{ansatz}) we then find that
\begin{equation}
     D_\mu \partial_a A_{\mu(3)} =  \hat {\bf r} \cdot 
     {\mbox{\boldmath $\tau$}} 
    \left[ \partial_a F' + {e\over 2} \partial_a F( H-2A + 3G) \right]
     = 0 .
\end{equation}
where the last equality follows from the variation of Eq.~(\ref{Fde})
together
with the relation $\partial_a G/G = 2\partial_a F/F $.   Thus, this mode
satisfies the background gauge condition without the need for any
additional gauge transformation, so
\begin{equation}
    \delta_a A_\mu = \partial_a A_\mu .
\end{equation}

    We can now use Eq.~(\ref{newmode}) to generate three other zero
modes from this mode.  Substitution of the expression (\ref{amode})
for $\delta_a A_\mu$ into this equation gives a mode that can be
written in the form\footnote{Showing that $\delta' A_{\mu(1)}$ and
$\delta' A_{\mu(3)}$ are of this 
form is trivial.  To verify the result for $\delta' A_{\mu(2)}$, one
must make use of the identity
     $( \partial_a L/ L )' =
       2 ( \partial_a F /F )' = 
      -  eG  ( \partial_a L / L )$
which is obtained by differentiating Eq.~(\ref{Fde}) with respect to
$a$.}
\begin{equation}
     \delta' A_\mu = D_\mu \Lambda = \partial_\mu \Lambda +ie[A_\mu,\Lambda] ,
\label{gaugeZ} 
\end{equation} 
where the only nonzero components of $\Lambda$ are 
\begin{equation}
  {\mbox{\boldmath $\Lambda$}}_{(2)}(r) =-\hat{\bf n}{\partial_a L \over eL} 
    = -{\hat {\bf n}\over e} \left[ {1\over a} - {1\over r} + O(r^{-2}) 
\right] . \label{bLambda}
\end{equation} 
This new mode is just a global $SU(2)$ zero mode, already in
background gauge.  Its relation to the gauge rotation angle is given
by $e\Lambda(\infty)$; from Eq.~(\ref{bLambda}), we see that 
the mode corresponding to a shift $\delta a$ maps to one
corresponding to an $SU(2)$ rotation by an angle $\delta \psi = \delta
a/a$.  

    The three translation zero modes are given by spatial derivatives
of the solution combined with appropriate gauge transformations.  Once
these are found, Eq.~(\ref{newmode}) can be used to obtain the eighth,
$U(1)$, mode.  We do not actually need the form of these four modes,
but we will make use of the fact that they are orthogonal to each
other and to the other four zero modes.  This orthogonality is clearly
expected on physical grounds.  To verify it, we first note that the
translation modes transform under spatial rotations as the components
of a vector, and so must be orthogonal to the other five modes, which
are rotational scalars.  It then follows that the Dirac mode from
which these arise is orthogonal to the Dirac mode obtained from the
$SU(2)$ and $\delta a$ modes; since the $U(1)$ mode arises from the
former Dirac mode, it must be orthogonal to the latter four modes.

\medskip
\subsection{The Moduli Space Metric}
\smallskip

     We can now proceed to determine the moduli space metric.
Symmetry considerations and the properties of the BPS mass formula
constrain its form considerably.  The subspace corresponding to the
translation modes is clearly $R^3$, with a natural set of coordinates
given by the location of the center of the monopole.  The metric on
this subspace relates the kinetic energy to the spatial velocity, and
so is proportional to the monopole mass, which depends only on the
magnetic charge.  Hence, it must be independent not only of the
position coordinates and $SU(2)$ and $U(1)$ parameters, but also of
the parameter $a$.  Similarly, since
the metric component $g_{\chi\chi}$ in the subspace spanned
by the $U(1)$ phase angle 
contributes to the leading corrections to the
dyon mass through a term of the form $Q_\chi^2/2 g_{\chi\chi}$, it too
must be independent of all eight parameters.
The subspace spanned
by the $SU(2)$ parameters must be simply the standard mapping of
$SU(2)$ onto a three-sphere, with a radius that might depend on $a$
but not on the position or $U(1)$
phase angle.  Finally, the metric in the one-dimensional $\delta a$
subspaces can depend at most on $a$.

    Thus the metric on the eight-dimensional moduli space
must be of the form 
\begin{equation}
    ds^2 = B d{\bf x}^2 + C d\chi^2 + I_1(a) da^2 
      + I_2(a) (\sigma_1^2 + \sigma_2^2 + \sigma_3^2) ,
\end{equation} 
where $B$ and $C$ are constants, and the one-forms $\sigma_j$ are
defined by 
\begin{eqnarray}
\sigma_1 &=& -\sin\psi d\theta +\cos\psi\sin\theta d\phi, \nonumber \\
\sigma_2 &=& \cos\psi d\theta +\sin\psi\sin\theta d\phi, \nonumber \\
\sigma_3 &=&  d\psi+\cos\theta d\phi\, ,
\end{eqnarray}
with the $SU(2)$ Euler angles $\theta$, $\phi$, and $\psi$ having 
periodicities $\pi$, $2\pi$, and $4\pi$, respectively.

    From Eq.~(\ref{modetometric}), we see that
$I_1(a)$ is simply the norm of the $\delta a$ mode; from its
construction, it is obvious that the $SU(2)$ mode of Eq.~(\ref{gaugeZ})
has the same norm.  Hence,
\begin{eqnarray}
     I_1(a) &=& \int d^3x \,{\rm tr}\,
         \left(\delta' A_\mu \delta' A_\mu \right) \nonumber\\
   &=& \int d^3x \,{\rm tr}\,
         \left(D_\mu\Lambda \right)^2 \nonumber\\
   &=& \int d^3x  \,\partial_j \left[ {\rm tr}\, (\Lambda D_j \Lambda)
              \right] \nonumber\\  
         &=& {4\pi \kappa \over e^2 a} ,
\end{eqnarray}
with $\kappa \equiv {\rm tr}\, t^3(\mbox{\boldmath $\gamma$})
t^3(\mbox{\boldmath $\gamma$})=1/\mbox{\boldmath $\gamma$}^2$.
In the second equality we have used the fact that $\delta' A_\mu$
obeys the background gauge condition, while in the last we have used
Eq.~(\ref{bLambda}).
To obtain $I_2(a)$ we need only multiply this by the 
square of the factor $\delta a /\delta \psi = a$ that followed from
$\Lambda(\infty)$.  Finally, $B$ and $C$ can be related to
the
monopole mass $M\equiv m_{\mbox{\boldmath $\beta$}}$ with the aid of 
the BPS dyon mass formula. 
We thus find that   
the moduli space metric is
\begin{equation}
    ds^2 = M d{\bf x}^2 +  {16\pi^2 \over M e^4}d\chi^2 + 
     {4 \pi \kappa \over e^2} \left[{ da^2 \over a} 
      + a (\sigma_1^2 + \sigma_2^2 + \sigma_3^2) \right] .
\label{firstform}
\end{equation} 
To put this in a more standard form, we define $\rho = 2\sqrt{a}$ and obtain
\begin{equation}
    ds^2 = M d{\bf x}^2 + {16\pi^2 \over M e^4} d\chi^2 + 
     \frac{4 \pi \kappa}{e^2} \left[ d\rho^2 
      + {\rho^2\over 4} (\sigma_1^2 + \sigma_2^2 + \sigma_3^2) \right] .
\end{equation} 
The quantity in square brackets is just the metric for $R^4$ written
in polar coordinates, with the unfamiliar factor of $1/4$ arising from
the normalization of the $\sigma_j$, and so the moduli space is 
\begin{equation}
    {\cal M} =  R^3 \times S^1 \times R^4 
\end{equation} 
with the natural flat metric.  (The second factor is $S^1$, rather than
$R^1$, because of the periodicity of $\chi$.)

We want to compare this with the metric for the moduli space of
solutions containing one ${\mbox{\boldmath $\beta$}}$- and one
${\mbox{\boldmath $\gamma$}}$-monopole in the MSB case. In
Ref.~\cite{klee2}, it was shown that this space is of the form
\begin{equation}
   {\cal M} = R^3 \times {R^1 \times {\cal M}_0 \over Z}.
\end{equation} 
Here ${\cal M}_0$ is the Taub-NUT space with metric
\begin{equation}
   {\cal G}_{{\cal M}_0} =
     \left(\mu +\frac{g^2\lambda}{8\pi r}\right)\,[dr^2+
r^2\sigma_1^2+r^2\sigma_2^2]+ \left(\frac{g^2\lambda}{8\pi}\right)^2
\left(\mu+\frac{g^2\lambda}{8\pi r}\right)^{-1} \sigma_3^2  ,
\end{equation} 
with the reduced mass $\mu=m_{\mbox{\boldmath $\beta$}} 
m_{\mbox{\boldmath $\gamma$}}/(m_{\mbox{\boldmath $\beta$}}+
m_{\mbox{\boldmath $\gamma$}})$ and the magnetic coupling $g=4\pi/e$.
The constant $\lambda$ encodes the strength of coupling between the
two monopoles,
\begin{equation}
\lambda=-2\mbox{\boldmath $\gamma$}^*\cdot\mbox{\boldmath $\beta$}^* 
    = 2 \kappa \, ,
\label{lambda}
\end{equation}
where the second equality follows from the fact that $\mbox{\boldmath
$\gamma$}$ is a long simple root of the non-simply-laced $SO(5)$
algebra.  The division by $Z$ denotes the fact that there is an
identification of points
\begin{equation}
   (\chi,\psi)=(\chi+2\pi,\psi+
   \frac{4m_{\mbox{\boldmath $\gamma$}}}{m_{\mbox{\boldmath $\beta$}}+
   m_{\mbox{\boldmath $\gamma$}}}\pi).
\label{identification}
\end{equation} 
Using Eq.~(\ref{lambda}) and the relation between $g$ and $e$, we see
that, as 
$\mu$ and $m_{\mbox{\boldmath $\gamma$}}$ tend to zero, ${\cal
G}_{{\cal M}_0}$ 
approaches the metric for the relative moduli space that we found for
the $SU(2)\times U(1)$ breaking, provided that we identify the radial
coordinate 
$r$ with the cloud size parameter $a$. 
Furthermore, in this limit the identification
(\ref{identification}) reduces to $(\chi,\psi) = (\chi +2\pi, \psi)$,
so the division by $Z$ acts only on the $R^1$ factor, allowing us to
make the replacement $R^1/Z = S^1$.  Thus the moduli space metric for
the NUS case is indeed the expected limit of that for the MSB case.

Although the metric varies smoothly as one case goes over into the
other, there is a curious transformation in the meaning of the
moduli space coordinates, specifically those on the four-dimensional
subspace that remains after the center-of-mass coordinates and overall
$U(1)$ phase have been factored out.  With maximal symmetry breaking
these coordinates are the distance $r$ between the 
${\mbox{\boldmath $\beta$}}$ and ${\mbox{\boldmath $\gamma$}}$ monopoles, 
the angles $\theta$ and $\phi$ that specify the
direction from one monopole to the other, and the relative $U(1)$
phase angle $\psi$.  As $\mu$ tends toward $0$ and the
${\mbox{\boldmath $\gamma$}}$-monopole ceases to be a distinct
soliton, the monopole separation $r$ becomes 
instead a measure of the size of the nonabelian cloud, 
while the directional angles $\theta$ and $\phi$ combine with $\psi$ to
give the coordinates in the internal symmetry space.

\subsection{Quantum Mechanics of the Moduli Space Coordinates}

In the moduli space approximation, one assumes that at
sufficiently low energy the classical dynamics of the monopoles is
mimicked by the free motion of a point particle on the moduli space.  
Quantizing this motion should then give the low energy quantum mechanics
of the monopoles.  This reduces the quantum mechanics of interacting
monopoles to a nonlinear sigma model with the moduli space as the
target manifold.  (When there are fermionic zero modes present, one
must modify the sigma model to include fermionic coordinates, but here
we want to confine our attention to the purely bosonic part.)

When the symmetry breaking is maximal, all bosonic coordinates on the
moduli space have a clear physical interpretation as either positions
or $U(1)$ phase angles of individual monopoles.  The periodicity of
the latter leads to the quantization of the dyonic charges.  On the
NUS moduli space of the $SO(5)$ solution found above, the
center-of-mass variables still have this interpretation.  Since the
corresponding portion of the moduli space is a flat $R^3\times S^1$, a
natural basis of energy eigenstates is given by plane waves on $R^3$
with a periodic dependence on the ``internal'' $S^1$; these describe a
freely propagating dyon with quantized electric $U(1)$ charge.

The relative part of this moduli space is a flat $R^4$, whose
coordinates may be taken as the cloud size parameter $a$ together with
$SU(2)$ gauge collective coordinates that span the transverse three-spheres.
$R^4$ admits an $SO(4)=SU(2)\times SU(2)$ isometry. Let us call the respective
$SU(2)$ generators $iL^{(a)}$ and $iK^{(a)}$, $a=1,2,3$. The wavefunction is 
then decomposed as
\begin{equation}
\Psi_{{\cal M}_0}=\sum A_{jlk}^Ef^{(j)}_E(a){\cal D}^{(j)}_{kl}
(\theta,\phi,\psi),
\end{equation}
where the ${\cal D}^{(j)}_{kl}$ are the three-dimensional spherical harmonics
that satisfy
\begin{eqnarray}
-L^{(a)}L^{(a)}\,{\cal D}^{(j)}_{kl}=-K^{(a)}K^{(a)}\,{\cal D}^{(j)}_{kl}
&=&j(j+1)\,{\cal D}^{(j)}_{kl}, \nonumber \\
iL^{(3)}{\cal D}^{(j)}_{kl}&=&l \,{\cal D}^{(j)}_{kl},\nonumber \\
iK^{(3)}{\cal D}^{(j)}_{kl}&=&k \,{\cal D}^{(j)}_{kl},
\end{eqnarray}
and $f^{(j)}_E(a)$ solves the eigenvalue equation 
\begin{equation}
-\frac{1}{a}\frac{d}{da} a^2\frac{d}{da}f^{(j)}_E+\frac{j(j+1)}{a}f^{(j)}_E=
Ef^{(j)}_E.
\label{Schrod}
\end{equation} 
As usual with representations of an $SU(2)$ group, the eigenvalues $l$ 
and $k$ are either integers or half-integers and are bounded by $-j$
and $j$. 

We will see in the next section that one triplet of generators,
$K^{(a)}$, induces $SU(2)$ global gauge transformations, so the
eigenvalue $j(j+1)$ encodes the electric $SU(2)$ (isospin) charge of
the resulting state.  The other triplet, $L^{(a)}$, are nothing
but the angular momentum in the center-of-mass frame.  Hence 
there is a tower of (non-BPS) states carrying both spin and isospin; 
the fact
that the eigenvalues of $L^{(a)}L^{(a)}$ are identical to those of
$K^{(a)}K^{(a)}$ implies that {\it the spin of the chromodyonic state is
identical to its isospin.}

This identity can be understood by considering the MSB case
first.  Both monopoles are then
massive and the angular momentum of the system is the sum of the
orbital angular momentum and an anomalous contribution, proportional
to the relative electric $U(1)$ charge $q$, of the form $q\hat{\bf
r}$.  Because these two contributions are orthogonal, $|q|$ gives a
lower bound on the magnitude of the total angular momentum that is
saturated when the orbital part vanishes.   
As the NUS limit is approached, the relative 
$U(1)$ is promoted to an $SU(2)$, so $|q|$ becomes the isospin. 
At the same time, one of the monopoles becomes massless and is
manifested as a spherically symmetric cloud about the other, so
the ``orbital'' angular momentum disappears.  The
equality of the isospin $|q|$ and the spin $j$ then follows. 

It is worth noting that this identity should hold beyond the BPS
limit.  Introducing a mass term for the Higgs scalar would lift the
degeneracy along the $a$-direction, so we would expect to find a
family of $SO(5)$ solutions similar to the above BPS solution but with
a definite size for the nonabelian cloud.  Because of the unbroken
$SU(2)$, the nonabelian gauge zero-modes would still span a
three-sphere in the appropriate moduli space and so should lead after
quantization to a tower of chromodyons with
the same eigenvalues for spin and isospin as before.

The quantization of the last collective coordinate, $a$, is less
transparent. Solving Eq.~(\ref{Schrod}) for the ground state (E=0)
radial wavefunction $f^{(0)}_{0}$, for instance, we find a unique
solution that is regular at the origin,
\begin{equation}
\Psi_{{\cal M}_0}(a)=constant,
\end{equation}
which is just the nonnormalizable, zero-momentum plane wave on the
$R^4$ with radial distance $\rho$.  In terms of the three-dimensional
monopole separation/cloud size parameter $a$, however, we have a
nontrivial probability distribution
\begin{equation}
|\Psi_{{\cal M}_0}|^2 \rho^3 d\rho \sim {1\over a} (a^2 da) \, .
\end{equation}
The proper physical interpretation of this result is just one of the
puzzles related to these states that we hope to investigate in the
future.

\section{The Symmetry of the Moduli Space}

In the previous section we showed in an $SO(5)$ example that the NUS
moduli space for a family of configurations carrying no net nonabelian
magnetic charge could be obtained as a limit of the known two-monopole
MSB moduli space.  More generally, the metric for the MSB moduli space 
was given in Ref.~\cite{klee} for all cases in which the monopoles are 
all fundamental and distinct.

\begin{center}
\leavevmode
\epsfysize=7.5in\epsfbox{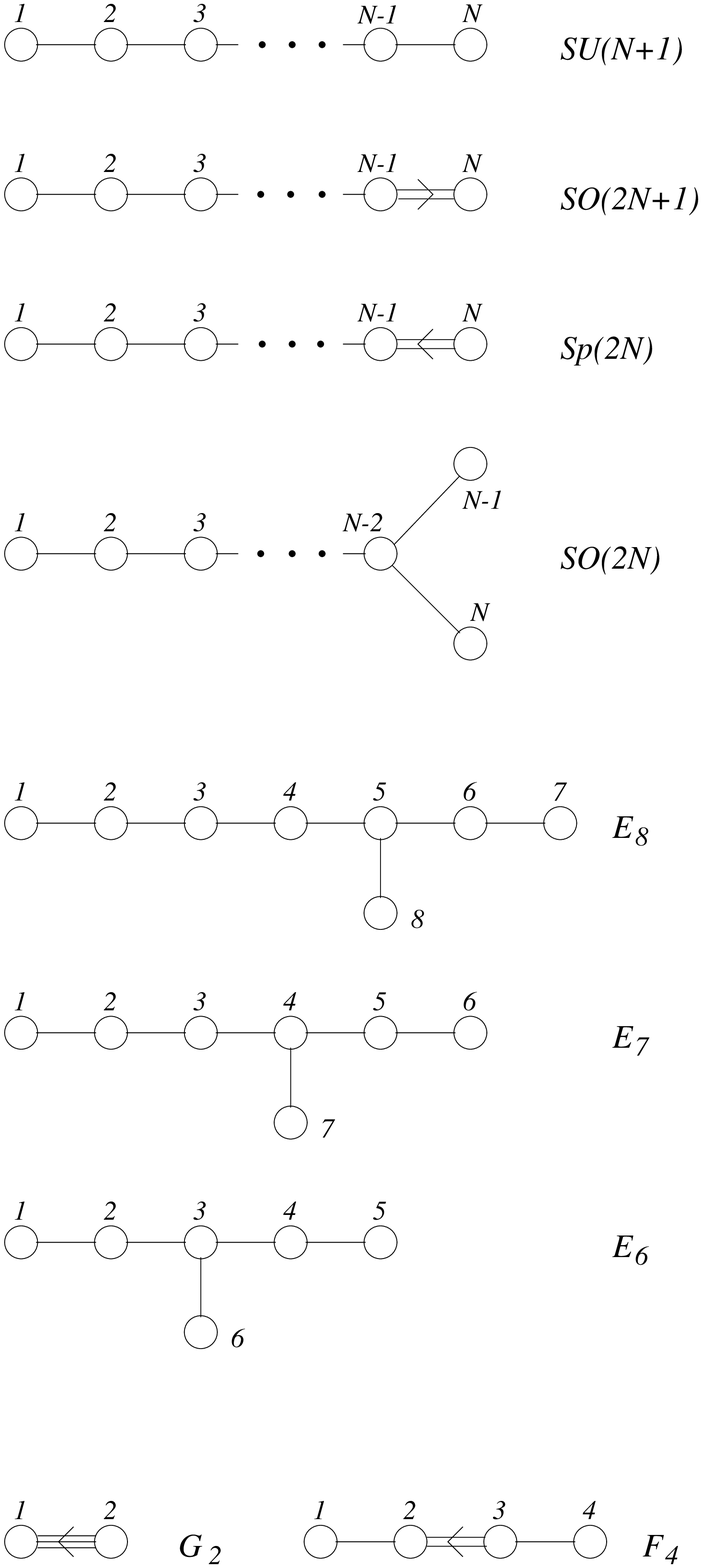}
\end{center}
\vskip 1cm
\begin{quote}
{\bf Figure 2:} {\small
Dynkin diagrams of all simple groups. For the non-simply-laced
cases, the arrow points toward the short roots. We  have also
labeled the simple roots for later reference.}
\end{quote}
\vskip 1cm

Before presenting this metric, we need some notation.
When a simple gauge group is maximally broken to its Cartan subgroup,
the fundamental monopoles are in one-to-one correspondence with the
$r$ simple roots $\mbox{\boldmath $\beta$}_a$, ($a=1,\dots,r$) of
the original gauge group.  A pair of such monopoles interact with each
other if and only if $\mbox{\boldmath $\beta$}_a \cdot \mbox{\boldmath
$\beta$}_b \neq 0$.  In the Dynkin diagram (see Fig.~2) such pairs are
indicated by linked circles. In any simple gauge group of rank $r$,
there are precisely $r-1$ such links.  We will label these links by an
index $A$, and denote by ${\bf r}_A$ the relative position vector
between the pair of fundamental monopoles connected by the $A$-th
link.  Likewise, $\psi_A$ is the linear combination of internal $U(1)$
angles that is conjugate to the relative $U(1)$ electric charge
between the two monopoles.  Finally, we generalize Eq.~(\ref{lambda}) by
defining $\lambda_A$ to be $-2$ times the inner product of the duals
of the roots joined by the $A$-th link.  
The relative part of the moduli space metric can then be written as
\begin{eqnarray}
{\cal G}_{\rm rel}&=&\sum_{A,B}C_{AB}d{\bf r}_A\cdot d{\bf r}_B \nonumber \\
&+&\sum_{A,B}\left(\frac{g^2}{8\pi}\right)^2\lambda_A\lambda_B(C^{-1})_{AB}
(d\psi_A+ {\bf w}_A\cdot d{\bf r}_A)(d\psi_B+{\bf w}_B\cdot d{\bf
r}_B) \, . 
\label{maximal}
\end{eqnarray}
Here the matrix $C$ is
\begin{equation}
C_{AB}=\mu_{AB}+\delta_{AB}\frac{g^2\lambda_A}{8\pi r_A},
\end{equation}
where $\mu_{AB}$ may be interpreted as a reduced-mass matrix,  and 
${\bf w}_A({\bf r}_A)$ is the vector potential due to a negative unit charged
Dirac monopole at ${\bf r}_A=0$.

In this section, we consider two types of configurations.  The first
is a direct generalization of the $SO(5)$ case, and consists of one
massive and $N$ massless monopoles in a theory with $Sp(2N+2)$ broken
to $U(1)\times Sp(2N)$.  Again the moduli space metric can be found by
direct calculation and then compared to the result obtained by the
limiting procedure. The second involves two massive and $N-1$ massless
monopoles in a theory with either $Sp(2N+2)$ or $SU(N+2)$ broken to
$U(1)^2\times SU(N)$.  In this case, a direct calculation of the
metric is not possible, since the full family of monopole solutions is
not known.  Although we cannot verify with certainty that the limiting
procedure yields the correct metric, we can check its consistency by
examining the symmetry of the moduli space metric. The unbroken gauge
symmetry must be realized as an isometry of the moduli space that
preserves the hyperk\"ahler structure, and the correct metric must
exhibit such properties. For both classes of theories, we give
explicit forms of the corresponding triholomorphic Killing vector
fields.

\subsection{Unbroken $Sp(2N)$}

The simplest class of examples arises when the gauge group 
$Sp(2N+2)$ is broken to $U(1)\times Sp(2N)$; the $SO(5)=Sp(4)
\rightarrow U(1)\times SU(2)$ example of the last section is a special
case of this.  We write the simple roots as  ${\mbox{\boldmath
$\beta$}}_1$ and ${\mbox{\boldmath $\gamma$}}_j$, ($j=2,\dots,N+1$),
with the indices corresponding to the numbering of roots in Fig.~2.
The sum
\begin{equation}
\frac{e{\bf g}}{4\pi}= {\mbox{\boldmath $\beta$}}_1^*+\sum_{j=2}^{N+1}
{\mbox{\boldmath $\gamma$}}_j^*
\end{equation}
is orthogonal to the ${\mbox{\boldmath $\gamma$}}_j$'s that span the
root lattice of the unbroken $Sp(2N)$, and gives the magnetic charge
of a configuration containing a single massive ${\mbox{\boldmath
$\beta$}}_1$ fundamental monopole surrounded by a  cloud of
massless monopoles that cancel the long-range nonabelian field.  

As mentioned above, this can be regarded as a generalization of the 
$SO(5)$ example of the previous section. In fact, we can identify an $SO(5)$
subgroup of $Sp(2N+2)$, generated by the pair ${\mbox{\boldmath $\gamma$}}
\equiv{\mbox{\boldmath $\gamma$}}_{N+1}$ and ${\mbox{\boldmath $\beta$}}\equiv
{\mbox{\boldmath $\beta$}}_1+\sum_{2}^{N}{\mbox{\boldmath
$\gamma$}}_j$, in which the 
$SO(5)$ solutions of the previous section can be embedded.
This embedding makes the Higgs expectation value $\bf h$ 
proportional to ${\mbox{\boldmath $\beta$}}^*+{\mbox{\boldmath $\gamma$}}^*=
{\mbox{\boldmath $\beta$}}_1^*+\sum_{j=2}^{N+1}{\mbox{\boldmath
$\gamma$}}_j^*$, which is just what is needed 
to ensure that the unbroken group is $Sp(2N)\times U(1)$.
Note that, even though the form of the
solution remains intact, the number of massless monopoles associated 
with this embedded solution is now $N$ rather than one. 

Further solutions can be obtained by gauge transforming such an
embedded solution by elements of the unbroken $Sp(2N)$, but not all
generators of $Sp(2N)$ transform it nontrivially. A generic embedded
solution is left invariant by $Sp(2N-2)$, and this tells us that there
must be at least ${\rm dim} [Sp(2N)/ Sp(2N-2)])=4N-1$ global gauge
zero modes. Since we already know that the $SO(5)$ solution contains
one parameter that fixes the size of the nonabelian cloud, the general
$Sp(2N+2)$ solution must admit at least one such  parameter.
Together, these account for all $4N$ coordinates of the relative
moduli space.  Let us now proceed to determine the metric of this space.

Consider a point on the moduli space corresponding to a generic
$SO(5)$ embedded solution. Since the geometry of the gauge orbit can
depend only on the parameter $a$, evaluating the metric at such a
point determines the metric everywhere.  Of the $4N-1$ gauge
generators that act on this point nontrivially, three arise from the
simple embedding and form an $SU(2)$, generated by $
{\bf t}(\mbox{\boldmath $\gamma$}_{N+1})$, that keep the solution within
the $SO(5)$ subgroup.  The other $4N-4$ gauge zero modes about such a
point are generated by the ladder
operators associated with the $2N-2$ positive roots ${\mbox{\boldmath
$\nu$}}_j= {\mbox{\boldmath $\gamma$}}_{j}+{\mbox{\boldmath
$\gamma$}}_{j+1}+\cdots+ {\mbox{\boldmath $\gamma$}}_{N}$ and
${\mbox{\boldmath $\mu$}}_j= {\mbox{\boldmath
$\gamma$}}_{j}+{\mbox{\boldmath $\gamma$}}_{j+1}+
\cdots+{\mbox{\boldmath $\gamma$}}_{N+1}$ with $j=2,\dots,N$.  The
associated zero modes $\delta A_\mu=D_\mu \Lambda$ satisfy
\begin{equation}
D_\mu D^\mu\Lambda=0,
\end{equation}
The solutions of this equation are found to be of the form 
\begin{equation}
\Lambda= \epsilon (r)\,T, \label{Tzero}
\end{equation}
where $T$ is any linear combination of the $4N-4$ ladder
operators above, appropriately normalized, and the radial function
$\epsilon$ satisfies
\begin{equation}
\frac{d\epsilon}{dr}+\frac{1}{2}G\epsilon=0, 
\end{equation}
where $G(r)$ is given by Eq.~(\ref{cloud}) and $\epsilon(\infty)=1/e$.

For generic values of $a$, the total gauge orbit must be topologically given 
by $Sp(2N)/Sp(2N-2)$ $= S^{4N-1}$, possibly up to a division
by a discrete  
group. Together with the fact that the last $4N-4$ gauge zero modes do not 
involve any of the $SU(2)$ generators $t^a(\mbox{\boldmath $\gamma$}_{N+1})$, 
this allows us to decompose the metric in the form
\begin{equation}
{\cal G}_{\rm rel}= \left[I_1(a)da^2+I_2(a)(\sigma_1^2+\sigma_2^2
+\sigma_3^2)+\sum_{s,t=4}^{4N-1}\tilde I_3^{st}(a)\sigma_s\sigma_t\right],
\end{equation}
where
\begin{eqnarray}
I_1(a)&=&\frac{4\pi\kappa}{e^2}\,\frac{1}{a}\nonumber \\
I_2(a)&=&\frac{4\pi\kappa}{e^2}\, a
\end{eqnarray}
were obtained in the previous section and
$\{\sigma_s/2, s=1,\dots,4N-1\}$ is an orthonormal frame on a
unit sphere $S^{4N-1}$.  

Further, the functional form of the gauge zero modes in Eq.~(\ref{Tzero})
is independent of the generator $T$, so we may choose $4N-4$ orthogonal
$T$'s, each of whose 
zero modes is given by Eq.~(\ref{Tzero}) with one and the same
function $\epsilon(r)$. Then,
\begin{equation}
\tilde I_3^{st}=I_3\,\delta_{st}\qquad 4\le s,t \le 4N-1,
\end{equation}
with 
\begin{equation}
I_3(a)=\int d^3x\;{\rm tr}\;D_\mu\Lambda D^\mu\Lambda=
\oint_\infty dS_\mu\;{\rm tr}\;\Lambda D^\mu\Lambda=
\frac{2\pi\kappa'}{e^2}a.
\end{equation}
Recall that $\kappa={\rm tr}\,t^3({\mbox{\boldmath $\gamma$}}_{N+1})
t^3({\mbox{\boldmath $\gamma$}}_{N+1})$, where the $SU(2)$ generators
$t^a({\mbox{\boldmath $\gamma$}}_{N+1})$ induce unit shifts along the
$\sigma_a$'s with $a=1,2,3$.  We must fix $\kappa'={\rm tr}\,T^2$ so
that $T$ will induce a unit shift along a $ \sigma_{s}$ with $4\le s\le
4N-1$.  The action of $Sp(2N)$ on an $S^{4N-1}$ is found by embedding
$Sp(2N)$ into $SO(4N)$; after normalizing all generators with respect
to the invariant bilinear form of $Sp(2N)$, we find that the
generators associated with the short roots $\mbox{\boldmath $\mu$}_j$
and $\mbox{\boldmath $\nu$}_j$ and those associated with the long root
${\mbox{\boldmath $\gamma$}}_{N+1}$ shift a point corresponding to a
generic $SO(5)$ embedding by unequal distances. The ratio turns out to
be $1/\sqrt 2$, so that the appropriate normalization for $T$ is such
that
\begin{equation}
\kappa'=2\kappa.
\end{equation}
The relative moduli space metric is then
\begin{equation}
{\cal G}_{\rm rel}=\frac{4\pi\kappa}{e^2}\left[\frac{1}{a}\,da^2+a
\sum_{s=1}^{4N-1}\sigma_s\sigma_s\right].
\end{equation}
After a coordinate redefinition $\rho=2\sqrt{a}$, we obtain
\begin{equation}
{\cal G}_{\rm rel}=\frac{4\pi\kappa}{e^2}\left[d\rho^2+\rho^2
\frac{\sum_{s=1}^{4N-1}\sigma_s\sigma_s}{4}\right]=
\frac{4\pi\kappa}{e^2}\left[d\rho^2+
\rho^2d\Omega^2_{4N-1}\right]=\frac{4\pi\kappa}{e^2}\, {\cal G}_{\rm
flat} \, ,
\label{NUSmetric}
\end{equation}
showing that the moduli space metric is that of a flat Euclidean
space $R^{4N}$. The smoothness of the metric at origin then requires the 
gauge orbit to be $S^{4N-1}$ globally, so  the relative moduli space is 
strictly $R^{4N}$.

To compare this to the NUS limit of the MSB metric in 
Eq.~(\ref{maximal}), let us first note that $\lambda_A=\lambda=
-2\mbox{\boldmath $\beta$}^*\cdot\mbox{\boldmath $\gamma$}^*$ for all
$A$.  We also need the fact that the $\mu_{AB}$ all vanish if there is
only a single massive monopole.  The NUS limit of  
Eq.~(\ref{maximal}) is then
\begin{equation}
 {\cal G}_{\rm rel} = \frac{g^2\lambda}{8\pi}\sum_{A}\left[\frac{1}{r_A}
d{\bf r}_A^2+ r_A(d\psi_A+\cos\theta_Ad\phi_A)^2\right],
\end{equation}
where we have rewritten the vector potential ${\bf w}_A\cdot d{\bf r}_A$ in 
polar coordinates.   
After a coordinate redefinition $\rho_A=2\sqrt{r_A}$, we see that this
MSB metric is a sum of $N$ copies of the flat $R^4$ metric,
\begin{equation}
{\cal G}_{\rm rel} =
\frac{g^2\lambda}{8\pi}\sum_{A}\left[d\rho^2+ \rho^2 \frac{\sigma_1^2+
\sigma_2^2+\sigma_3^2}{4}\right]_A =\frac{g^2\lambda}{8\pi}\sum_{A}
\left[d\rho^2+ \rho^2 d\Omega_3^2\right]_A
=\frac{g^2\lambda}{8\pi}\,{\cal G}_{\rm flat}.
\end{equation}
Since $\lambda/2=\kappa$ and $eg=4\pi$, this is the same as the metric of 
Eq.~(\ref{NUSmetric}), thus verifying that the two approaches 
produce the same result.

It is curious that the single $Sp(2N)$-invariant $a$ can
be written as the sum of all distances between adjacent (as defined by
the Dynkin diagram) monopoles, massive or massless alike; i.e.,
\begin{equation}
  a =\frac{1}{4}\rho^2  =\frac{1}{4}\sum_A\rho_A^2=\sum_A r_A.
\end{equation}
However, the fact that there is only a single invariant parameter
implies that the individual $r_A$'s are not invariant (the results of
the next section will make this more explicit). Thus, the positions of
the massless monopoles do not have a gauge-invariant meaning,
emphasizing again that the massless monopoles should not be regarded as
localized objects.

As for the $SO(5)$ example of Sec.~3, a tower of non-BPS states
carrying nonabelian electric charge can be constructed by
semiclassical quantization of the moduli space coordinates.  The
degeneracy of these states will be greater than in that example,
reflecting the  greater symmetry of the higher-dimensional moduli
space.

\subsection{The $Sp(2N)$ Triholomorphic Isometry of $R^{4N}$}

Because $Sp(2N)$ is an unbroken symmetry of the field equations and of
the boundary conditions, its action on the solutions will manifest
itself as metric-preserving diffeomorphisms of the relative moduli
space $R^{4N}$.  In fact, since the relative moduli space is a flat
$R^{4N}$, it possesses a larger isometry group, $SO(4N)$. However, the
$Sp(2N)$ subgroup acquires a special significance because it is the
maximal subgroup of $SO(4N)$ that preserves the hyperk\"ahler
structure of the manifold. This triholomorphicity is a generic feature
of isometries associated with the gauge rotation.

Let us introduce a pair of complex coordinates $\xi_A = 
x^1_A+ix^2_A$ and $\zeta_A=x^3_A+ix^4_A$
in the $R^4$ spanned by ${\bf r}_A$ and $\psi_A$. These are related to the 
relative Euler angles and the monopole separation $r_A$ by
\begin{eqnarray}
\xi_A&=&2\sqrt{r_A}\cos(\theta_A/2)e^{-i\,(\phi_A+\psi_A)/2},\nonumber\\
\zeta_A&=&2\sqrt{r_A}\sin(\theta_A/2)e^{-i\,(\phi_A-\psi_A)/2}.\label{43}
\end{eqnarray}
In effect, we have chosen a particular complex structure on $R^{4N}$. 
Given this complex structure, the k\"ahler form  is
\begin{eqnarray}
w^{(3)}&=&\frac{i}{2}\sum_A(d\xi_A \wedge d\xi_A^*+d\zeta_A \wedge
d\zeta_A^*) \nonumber \\
  &=&-\sum_A \left(\frac{1}{r_A}dr_A^1\wedge dr_A^2+dr_A^3\wedge
(d\psi_A+\cos\theta_A d\phi_A)\right) \, ,
\end{eqnarray}
where in the second line we have used 
\begin{eqnarray}
r^1_A-ir_A^2&=&\frac{\zeta_A\xi_A}{2} ,\nonumber\\
r^3_A&=& \frac{\xi_A \xi_A^*-\zeta_A^*\zeta_A}{4} .\label{cartesian}
\end{eqnarray}

Each factor of $R^4$ admits an $SO(4)=SU(2)\times SU(2)$ isometry. The first
$SU(2)$ is generated by
\begin{eqnarray}
L_A^{(3)}&=&\frac{i}{2}\left(\xi_A\frac{\partial}{\partial \xi_A}+
\zeta_A\frac{\partial}{\partial \zeta_A}-\xi_A^*\frac{\partial}{\partial 
\xi_A^*}-\zeta_A^*\frac{\partial}{\partial \zeta_A^*}\right)
=-\frac{\partial}{\partial\phi_A}\nonumber \\
L_A^{(+)}&=&i\left(\zeta_A^*\frac{\partial}{\partial \xi_A}-
\xi_A^*\frac{\partial}{\partial \zeta_A}\right)\nonumber\\
L_A^{(-)}&=&i\left(\xi_A\frac{\partial}{\partial \zeta_A^*}
-\zeta_A\frac{\partial}{\partial \xi_A^*}\right).
\end{eqnarray}
After re-expressing these in terms of ${\bf r}_A$ and $\psi_A$, and
writing 
$L_A^{(\pm)}=L_A^{(1)}\pm i L_A^{(2)}$, we can add these to obtain  
\begin{equation}
{\bf L}= \sum_A {\bf L}_A =
 \sum_A\left[  -{\bf r}_A\times\left({\bf \nabla}_A-{\bf w}_A
\frac{\partial}{\partial \psi_A}\right)-\hat{\bf r}_A\frac{
\partial}{\partial \psi_A} \right], \label{rotation}
\end{equation}
which is the standard form for the generators of three-dimensional
rotations in the presence of the vector potentials ${\bf w}_A$.  Under
appropriate rotations induced by $\bf L$,
the k\"ahler two-form $w^{(3)}$ is transformed into the other two
k\"ahler forms
\begin{equation}
w^{(a)}=w^{(a)}_{\rm flat}\equiv
-\sum_A \left(\frac{1}{2r_A}\epsilon^{abc}dr_A^b\wedge dr_A^c+
dr_A^a\wedge (d\psi_A+\cos\theta_A d\phi_A)\right),
\end{equation}
that are needed to complete the hyperk\"ahler structure of the moduli space.

In contrast, the second $SU(2)$ is holomorphic and thus cannot rotate the
complex structure. Its generators are given by
\begin{eqnarray}
K_A^{(3)}
&=&\frac{i}{2}\left(-\xi_A\frac{\partial}{\partial \xi_A}+
\zeta_A\frac{\partial}{\partial \zeta_A}+\xi_A^*\frac{\partial}{\partial 
\xi_A^*}-\zeta_A^*\frac{\partial}{\partial \zeta_A^*}\right) 
=\frac{\partial}{\partial \psi_A} ,\nonumber\\
K_A^{(+)}&=&i\left( \xi_A\frac{\partial}{\partial \zeta_A}
           -\zeta_A^*\frac{\partial}{\partial \xi_A^*}\right),\nonumber\\
K_A^{(-)}&=&i\left(\zeta_A\frac{\partial}{\partial \xi_A}- 
           \xi_A^*\frac{\partial}{\partial \zeta_A^*}\right).
\end{eqnarray}
Since these $K_A^{(a)}$'s commute with the $L_A^{(a)}$'s and since 
$\bf L$ induces rotations among the $w^{(a)}$'s, 
the $K_A^{(a)}$'s are in fact triholomorphic; i.e., they
preserve the hyperk\"ahler structure of the moduli space:
\begin{equation}
{\cal L}_{K_A^{(a)}}w^{(a')}_{\rm flat}=0.
\end{equation}
These $SU(2)$'s are clearly part of the $Sp(2N)$ isometry, with the
$K_A^{(3)}$'s forming a set of $N$ commuting generators that can be
taken to be the generators of the Cartan subalgebra.

To complete the $Sp(2N)$, we recall that $Sp(2N)$ contains an
$SU(N)\times SU(2)$ subgroup. Let the $SU(N)$ be generated by the
simple roots $\{{\mbox{\boldmath $\gamma$}}_2,\dots,{\mbox{\boldmath
$\gamma$}}_{N}\}$ and the $SU(2)$ by $\{{\mbox{\boldmath
$\gamma$}}_{N+1}\}$. Since this $SU(2)$ maps a given $SO(5)$ embedded
solution to another embedded solution in the same $SO(5)$ subgroup, it
must be realized on 
the moduli space by the $K_N^{(a)}$'s, which form the unique
triholomorphic $SU(2)$ that preserves the 4-plane $r_1 = r_2 = \dots =
r_{N-1} =0$.  The $SU(N)$, on the other hand, rotates one $R^4$ to
another; its Killing vectors are
\begin{eqnarray}
T_A&=&\frac{1}{2}\left( K_A^{(3)}-K_{A+1}^{(3)}\right), \nonumber\\
E_{AB}&=&\frac{i}{\sqrt 2}\left( \xi_A\frac{\partial}{\partial \xi_B}
                    +\zeta_A^*\frac{\partial}{\partial \zeta_B^*}
                    -\xi_B^*\frac{\partial}{\partial \xi_A^*}
                    -\zeta_B\frac{\partial}{\partial \zeta_A}\right),
\qquad A\neq B. \label{SUn}
\end{eqnarray}
Commuting the $E_{AB}$'s with the $K^{(\pm)}_{B}$'s results in another
set of ladder operators, 
\begin{eqnarray}
\tilde E_{AB}^{(+)}&=&\frac{i}{\sqrt 2}
                    \left( \xi_A\frac{\partial}{\partial \zeta_B}
                    +\xi_B\frac{\partial}{\partial \zeta_A}
                    -\zeta_A^*\frac{\partial}{\partial \xi_B^*}
                    -\zeta_B^*\frac{\partial}{\partial \xi_A^*}\right),
\qquad A<B, \nonumber \\
\tilde E_{AB}^{(-)}&=&\frac{i}{\sqrt 2}
                    \left( \zeta_A\frac{\partial}{\partial \xi_B}
                    +\zeta_B\frac{\partial}{\partial \xi_A}
                    -\xi_A^*\frac{\partial}{\partial \zeta_B^*}
                    -\xi_B^*\frac{\partial}{\partial \zeta_A^*}\right),
\qquad A<B,
\end{eqnarray}
that give the remaining generators of $Sp(2N)$.  These all commute
with $\bf L$ and are therefore triholomorphic.  We thus
have the required triholomorphic isometry,
\begin{equation}
\{K^{(3)}_A, E_{AB}, \tilde E_{AB}^{(\pm)}, K_A^{(\pm)},
1\le A\neq B\le N\} \quad\rightarrow \quad  Sp(2N).
\end{equation}

These generators have particularly simple interpretations in terms of
an orthonormal basis ${\bf e}_j$ ($j= 1, \dots, N$) in the root space of
$Sp(2N)$. (In this basis, ${\mbox{\boldmath $\gamma$}}_{j+1}= ({\bf e}_{j} 
- {\bf e}_{j+1})/2$ for $j<N$, while ${\mbox{\boldmath $\gamma$}}_{N+1} = 
{\bf e}_{N}$.)  The correspondence between roots and ladder operators is 
then simply
\begin{eqnarray}
E_{AB} &\rightarrow& \frac{1}{2}({\bf e}_A - {\bf e}_B),\nonumber \\
\tilde E_{AB}^{(\pm)} &\rightarrow&  \pm\frac{1}{2}({\bf e}_A +{\bf e}_B), 
\nonumber \\
K_A^{(\pm)} &\rightarrow&  \pm {\bf e}_A .
\end{eqnarray}

Finally, we note that in terms of the complex coordinates, 
the invariant $a$ takes the simple form
\begin{equation}
a=\sum_A r_A= {1\over 4}\sum_A (\xi_A\xi_A^*+\zeta_A^* \zeta_A) \, ,
\end{equation}
which is manifestly invariant under the transformations generated by
the Killing vectors.

\subsection{Unbroken $SU(N)$}

A slightly more involved example arises when either $Sp(2N+2)$ or
$SU(N+2)$ is broken to $SU(N)\times U(1)^2$. One finds that the magnetic 
charge 
\begin{equation}
\frac{e{\bf g}}{4\pi}={\mbox{\boldmath $\beta$}}_1^*+\sum_{j=2}^{N}
{\mbox{\boldmath $\gamma$}}_j^*+{\mbox{\boldmath $\beta$}}_{N+1}^*
\label{single}
\end{equation}
(where the roots are again numbered in accordance with the Dynkin
diagram of Fig.~2) is orthogonal to the ${\mbox{\boldmath
$\gamma$}}_j$'s that span the unbroken $SU(N)$.  This corresponds to a
combination of two massive monopoles, associated with
${\mbox{\boldmath $\beta$}}_1^*$ and ${\mbox{\boldmath $\beta$}}_{N+1}^*$ 
and having masses $m_1$ and $m_{N+1}$, and $N-1$ massless monopoles.

The relative moduli space again has the topology of $R^{4N}$ and can
be covered by the coordinate system $\{\xi_A,\zeta_A\}$ defined in the
previous subsection.  However, this moduli space is no longer flat.
Referring to the results of Ref.~\cite{klee2}, one finds that the reduced 
mass matrix $\mu_{AB}$ no longer vanishes.  Instead, the $N^2$ elements of
this matrix are all equal to the reduced mass of the two massive
monopoles; i.e., 
\begin{equation}
\mu_{AB}=\bar\mu\equiv\frac{m_1 m_{N+1}}{m_1+m_{N+1}}\, , \quad \hbox{all $A$
and $B$}.
\end{equation}
Using this and the fact that $\lambda_A=\lambda$ is again independent of 
the link index $A$, we find that the NUS limit of the MSB 
metric is
\begin{equation}
{\cal G}_{\rm rel}=\frac{g^2\lambda}{8\pi}
{\cal G}_{\rm flat}+\bar\mu\left(\sum_A d{\bf r}_A \right)^2-
\frac{g^2\lambda\bar\mu}{g^2\lambda+8\pi\bar\mu\sum_B r_B}
\left(\sum_A r_A\,(d\psi_A+\cos\theta_Ad\phi_A)\right)^2. \label{metric}
\end{equation}
The metric is still hyperk\"ahler, as it must be, but the 
three independent k\"ahler forms are now given by \cite{4N,gary}
\begin{eqnarray}
w^{(a)}_{SU(N)}&=&-\frac{1}{2}\sum_{A,B} C_{AB}\epsilon^{abc}dr_A^b
\wedge dr_B^c -\frac{g^2\lambda}{8\pi}\sum_A dr_A^a\wedge (d\psi_A+
\cos\theta_A d\phi_A)\nonumber\\
&=&\frac{g^2\lambda}{8\pi}w^{(a)}_{\rm  flat}-\frac{\bar \mu}{2}
\epsilon^{abc}\left(\sum_A dr^b_A\right)\wedge \left(\sum_B dr_B^c\right).
\label{3k}
\end{eqnarray}

As in the previous example, this moduli space must reflect the
symmetries of the theory.  There must be three Killing vector fields
that generate three-dimensional rotations of the multimonopole
solution, while the unbroken gauge symmetry must be realized as a 
triholomorphic $U(N)$ isometry with appropriate Killing vectors.  Now
note that if the original gauge group is $Sp(2N+2)$, this example
reduces to our previous one in the limit $\bar\mu \rightarrow 0$.
Hence, in that limit the rotational Killing vectors for the present
case must reduce to the $\bf L$ of Eq.~(\ref{rotation}), while the
triholomorphic Killing vectors must reduce to those that generate the
$U(N)$ subgroup of $Sp(2N)$ in Eq.~(\ref{SUn}).

In fact, the vector fields $\bf L$ in Eq.~(\ref{rotation}) generate the 
rotational $SU(2)$ isometry of the general MSB metric in Eq.~(\ref{maximal}) 
\cite{klee2}, and thus are Killing vectors on the NUS moduli space as well.
Further, it turns out, as we show in the Appendix, that the $SU(N)$
generators in Eq.~(\ref{SUn}) are  
Killing vectors on this curved moduli space and also preserve the 
hyperk\"ahler structure. Together with a simultaneous rotation of all 
$\psi_A$, this completes the $U(N)$ triholomorphic isometry induced by 
the action of the unbroken gauge group on the multimonopole solutions,
\begin{eqnarray}
\{ T_A, E_{AB},1\le A\neq B\le N\}&\rightarrow& SU(N), \nonumber \\
K\equiv \sum_A K_A^{(3)}&\rightarrow& U(1) .
\end{eqnarray}
Physically $K$ corresponds to the relative electric $U(1)$ charge of the 
two massive monopoles.
This $U(N)$ contains the $N$ $U(1)$ generators $K_A^{(3)}=\partial/
\partial\psi_A$ that clearly preserve both the metric and the 
hyperk\"ahler structures of the general MSB metric. This is not true
in general for the $E_{AB}$'s, but the detailed calculation in the Appendix
shows that they are all triholomorphic and metric-preserving in the
present NUS limit.
An important consistency check is to see if these vector fields preserve 
such gauge-invariant quantities as the relative position 
vector and the relative $U(1)$ charge of the two massive monopoles.\footnote{
In fact this is sufficient to show that $E_{AB}$ generates a symmetry of
the monopole dynamics. Consider the first order form of the Lagrangian
\begin{eqnarray}
L(\bar \mu)&=& \frac{1}{2}\sum_{A,B}C_{AB}\left[\dot{\bf r}_A\cdot
\dot{\bf r}_B-
q_Aq_B\right]+ \sum_{A}q_A(\dot\psi_A+\cos\theta_A\dot\phi_A)\nonumber\\
&=& L(\bar\mu=0)+
\frac{\bar \mu}{2}\left[\left(\sum_A \dot{\bf r}_A\right)^2-\left(
\sum_A q_A\right)^2\right].
\end{eqnarray}
The metric is recovered by integrating out the conserved charges $q_A$ 
(which is conjugate to $\psi_A$) and replacing velocities by line elements. 
$L(\bar \mu=0)$ is by itself invariant since it describes free motion on 
$R^{4N}$. The invariance of the relative position $\sum_A{\bf r}_A$
and the relative $U(1)$ charge $\sum q_A$ then implies the invariance of
the whole Lagrangian.}
The latter is clearly invariant since its charge operator $-iK$ 
commutes with all generators of $SU(N)$, while the former is also 
invariant if
\begin{equation}
E_{AB}\left[ \sum_A {\bf r}_A\right]=0 .
\end{equation}
One can show this explicitly using Eq.~(\ref{cartesian}).

The $4N$ coordinates of the moduli space can be related to
the physical parameters of the multimonopole solution.  An embedding
argument similar to that of Sec.~4.1 shows that the gauge orbit of a
generic point in the moduli space is of the form $U(N)/U(N-2)$. This
implies that the number of gauge modes is $4N-4$, including one that
corresponds to the relative $U(1)$ phase of the two massive monopoles.
Three more parameters must correspond to the relative position vector
between the two massive monopoles.  This leaves only one
gauge-invariant coordinate to characterize the nonabelian cloud.  A
natural choice for this coordinate is just 
$a=\sum_A r_A$, 
which we know from our previous results to be $U(N)$-invariant.  
In the simplest case,  
with $N=2$, the gauge orbits are ellipsoids, $a=r_1+r_2=const$, with focal 
points at the two massive monopoles; these become three-spheres if the two
massive monopoles coincide.

As with our previous examples, the semiclassical quantization of 
the moduli space coordinates will lead to a tower of chromodyonic
states.  A new feature here is that the relevant ``moment of inertia''
will increase with the separation between the two massive monopoles,
in a fashion similar to that found in Ref.~\cite{nelson}.

\section{Characteristics of General Moduli Spaces}

In the examples of the previous two sections we were able to analyze
monopoles in theories with unbroken nonabelian symmetries by taking
the appropriate limit of the MSB case.  For combinations of monopoles
such that the long-range magnetic field was invariant under the action of
the unbroken nonabelian symmetry, the NUS limit of the MSB moduli
space was shown to possess an isometry corresponding to the unbroken
gauge symmetry.  We found that color-magnetically neutral
combinations were composed of a number of massive monopoles surrounded
by a nonabelian cloud that could be viewed as arising from the
coalescence of a number of massless monopoles carrying purely
nonabelian magnetic charges.  In this section we will consider the
general case of a simple group $G$ of rank $r$ broken to $K\times
U(1)^{r-k}$ with a simple group $K$ of rank $k$.  (The extension to
semisimple $K$ is straightforward.)  As before, we denote the simple
roots of $K$ by ${\mbox{\boldmath $\gamma$}}_i$ and the remaining
roots of $G$ by ${\mbox{\boldmath $\beta$}}_a$, with numbering
corresponding to the Dynkin diagrams of Fig.~2.  

Although we do not know the moduli space metric for most cases, we can
still learn a good deal about how the massive and massless monopoles
combine to form neutral configurations.  In such configurations the
magnetic charge vector $\bf g$ must be orthogonal to every root of
$K$; i.e., 
\begin{eqnarray}
0 & = & \frac{e{\bf g}}{4\pi}\cdot {\mbox{\boldmath $\gamma$}}_j  \nonumber\\
  & = &   \sum_{a=1}^{r-k}n_a{\mbox{\boldmath $\beta$}}_a^*
\cdot{\mbox{\boldmath $\gamma$}}_j + \sum_{i=1}^{k}q_i{\mbox{\boldmath 
$\gamma$}}_i^*\cdot{\mbox{\boldmath $\gamma$}}_j .
\label{gmono}
\end{eqnarray}
for all $j$.  The sum of any two or more such ${\bf g}$'s will also
satisfy this condition; we will concentrate here on the `minimal'
cases, for which {\bf g} cannot be decomposed as such a sum.

The number of normalizable zero modes about such solutions, i.e.  the
dimension of the moduli space ${\cal M}$, is equal to $4(n+q)$, where
$n=\sum n_a$ and $q=\sum q_i$ are the number of massive and massless
fundamental monopoles, respectively \cite{erick2}.  The examples
described above suggest that $4n$ of these describe the position
coordinates and $U(1)$ phases of the massive monopoles, while the
remaining $4q$ describe the nonabelian cloud.  Of the latter, some
describe the size and, possibly, other gauge-invariant characteristics
of the nonabelian cloud and the rest correspond to global nonabelian
gauge rotations of the configuration.  The number of such gauge modes
can be as large as the dimension of $K$, but is less if the generic
solution is invariant under some group $K'\subset K$.  (Note that $K'$
need not be semisimple.) The number of parameters describing the
gauge-invariant structure of the cloud is then
\begin{equation}
{\cal N}_{\rm structure} = 4q- {\rm dim}[K/K'].
\label{shape}
\end{equation}

The problem of finding the minimal ${\bf g}$'s can be phrased in terms
of group representations. Each of the massive monopoles
transforms according to a representation of the dual group $K_{dual}$
spanned by the ${\mbox{\boldmath $\gamma$}}_j^*$.  (The dual group
enters here because the magnetic charge $\mbox{\boldmath $\beta$}_a^*$
is a weight vector with respect to the dual root system spanned by the
${\mbox{\boldmath $\gamma$}}_j^*$ \cite{topology}.)  The desired $\bf
g$'s correspond 
to collections of massive monopoles that can be combined with a number
of adjoint representation massless monopoles to form a group
singlet.\footnote{An equivalent approach starts from the 
observation that any fundamental weight is a linear combination of the
simple roots with nonnegative rational coefficients.  By definition, a
fundamental weight is orthogonal to all but one simple root, so any
linear combination of the fundamental weights associated with the
broken simple roots $\mbox{\boldmath $\beta$}_a^*$'s of $G_{dual}$ is
automatically orthogonal to the $\mbox{\boldmath $\gamma$}_j^*$'s and
hence to the $\mbox{\boldmath $\gamma$}_j$'s.  To obtain a minimal
$\bf g$, one simply adds a number of such fundamental weights in such
a way that the coefficients of the $\mbox{\boldmath $\beta$}_a^*$'s
and of the $\mbox{\boldmath $\gamma$}_j^*$'s in the final expression
are all nonnegative integers; these coefficients then give the number
of massive and massless monopoles required for the configuration. }

The representations of the massive fundamental monopoles can be
identified with the aid of the Dynkin diagram.  Consider the monopole
corresponding to the root $\mbox{\boldmath $\beta$}_a$ and let
$\mbox{\boldmath $\gamma$}_j$ be the root of $K$ to which
$\mbox{\boldmath $\beta$}_a$ is linked in the Dynkin diagram.  (If
$\mbox{\boldmath $\beta$}_a$ is not linked to a root of $K$, then the
monopole transforms as a singlet.)  If $\lambda \equiv -2
\mbox{\boldmath $\beta$}_a^* \cdot \mbox{\boldmath $\gamma$}_j^* = 1$,
then the monopole transforms according to the complex conjugate of
the basic representation of
$K_{\rm dual}$ corresponding to $\mbox{\boldmath $\gamma$}_j^*$; if
$\lambda > 1$, then the monopole transforms as a symmetric product of
$\lambda$ such representations.

With these ideas in mind, let us recall the case of $Sp(2N+2)
\rightarrow Sp(2N) \times U(1)$ that was considered in Sec.~4.1.  The
single massive monopole is linked to the first root of $Sp(2N)$, with
$\lambda=1$. It therefore transforms according to the vector
representation of the dual group, $K_{dual}=SO(2N+1)$.  Since the adjoint
representation of an orthogonal group is the antisymmetric product of
two vectors, and the antisymmetric product of $2N+1$ vectors is a
singlet, a color-neutral combination can be obtained by combining the
massive monopole with $N$ massless monopoles, in agreement with our
previous results.

In the other case considered in Sec,~4, with $K=K_{\rm dual}=SU(N)$
and $G$ being either $SU(N+2)$ or $Sp(2N+2)$, the two massive
monopoles were linked to the first and last simple roots of the
unbroken $SU(N)$, both with $\lambda=1$.  These therefore transform
under the defining representations $F$ and $\bar F$.  The neutral
combination of Eq.~(\ref{single}) corresponds to the fact that a group
singlet can be formed by combining an $F$ and an $\bar F$ with a
number of adjoints.  However, this is not the only possibility.  A
singlet can also be constructed from $N$ $F$'s (or $N$ $\bar F$'s)
together with some adjoint representation objects.  The corresponding
color-neutral magnetic charge is
\begin{equation}
\frac{e{\bf g}}{4\pi} =  N{\mbox{\boldmath $\beta$}}_1^*+ 
\sum_{j=1}^{j=N-1}(N-j){\mbox{\boldmath $\gamma$}}_{j+1}^*  .
\end{equation}
This describes a family of solutions composed of $N$ massive and
$N(N-1)/2$ massless monopoles, with a moduli space of dimension
$2N^2+2N$. The positions and $U(1)$ phases of the $N$ massive
monopoles account for $4N$ of these. There appears to be no invariance
subgroup, so there are $N^2-1$ gauge modes from the global $SU(N)$
rotations.  This leaves $(N-1)^2$ structure parameters that encode the gauge
invariant characteristics of the nonabelian cloud; this shows that the
cloud can have much more structure than it did in our $SO(5)$ example.

With other choices for $G$, additional representations of $SU(N)$ can
arise.  If $G=SO(2N+1)$, one can have a massive
monopole linked to the last simple root of $SU(N)$ with $\lambda=2$,
corresponding to the symmetric rank two tensor representation $S$,
while with $G=SO(2N)$ a massive monopole can be linked to the next to
last root of $SU(N)$, with $\lambda=1$, giving an antisymmetric rank
two tensor $\Lambda$.  In addition, the even orthogonal groups allow
the possibility of two different monopoles transforming as
fundamentals; this can happen if the last two simple roots of $SO(2N)$
are broken but ${\mbox{\boldmath $\gamma$}}_{N-2}$ is not.  A few more
possibilities arise for low values of $N$ by taking $G$ to be an
exceptional group.  An antisymmetric rank three tensor representation
$\Delta$ can be obtained if $G=E_6$, $E_7$, or $E_8$, while there is a
breaking of $G_2$ to $SU(2)\times U(1)$ that gives monopoles
transforming under the spin 3/2 representation of $SU(2)$.

For $K= SO(2N+1)$, $ SO(2N)$, or $Sp(2N)$, there is one type of
massive monopole, transforming under the defining or vector
representation $V$, if $G$ is a classical group.  If $G$ is
exceptional, there can also be massive monopoles transforming under the
spinor representations corresponding to the last root of $SO(2N+1)$
and the last two roots of $SO(2N)$ or under the 14-dimensional
representation corresponding to the last root of Sp(6).  

In Table~1 we list, for the case where the original gauge group is a
classical group, the various ways in which these representations can
be combined to give minimal configurations with vanishing nonabelian
magnetic charge.  The overall group $G$ that is shown is the smallest
one that allows the neutral combination shown; in most cases a larger
$G$ is also possible.  In the table we also give the decomposition of
$\bf g$ into simple roots, with the coefficients corresponding to
massive monopoles indicated by boldface type.  The remaining
coefficients give the number of massless monopoles, from which in turn
the total number of gauge and cloud structure zero modes can be obtained.

For the neutral combinations in which the number of component massive
monopoles is independent of the rank of the group, the number of massless
monopoles grows linearly with $N$. Since the dimension of $K$ grows
quadratically with the rank, there must be a nontrivial invariance subgroup
$K'$.  In all such cases, the generic solution for sufficiently high rank
can be obtained by an embedding of a lower rank solution.  Thus, the
solutions studied in Sec.~4.1 for $Sp(2N+2)$ broken to $Sp(2N)$ could all
be obtained by embedding the $SO(5)$ solution, and had $K'=Sp(2N-2)$, while
the solutions with two massive monopoles considered in  Sec.~4.3 were all
equivalent to embeddings either of $SU(4)$ solutions [if $G=SU(N+2)$] or of
$Sp(6)$ solutions [if $G=Sp(2N+2)$], and had $K'=U(N-2)$.  

\vskip 1cm
\hskip -6mm
{\small
\begin{tabular}{|l|l|l|l|l|l|} \hline
$K$      
          & Singlet     
                   & $G$    
                              & $e{\bf g}/4\pi$   
                                         & $K'$ 
                                                    & ${\cal N}_{\rm
structure}$ 
\\ \hline
$SU(N)$
         &  $\bar F^N$ 
                   & $SU(N+1)$ 
                             & $({\bf N},N-1,\dots,2,1)$ 
                                         &.         & $(N-1)^2$ \\
\cline{2-6}
         &  $F^N$ 
                   & $Sp(2N)$ 
                             & $(1,2,\dots,N-1,{\bf N})$ 
                                         &.         & $(N-1)^2$ 
\\  \cline{2-6}
         & $F\bar{F}$ 
                   & $SU(N+2)$ 
                             & $({\bf 1},1,\dots,1,{\bf 1})$
                                         & $U(N-2)$ 
                                                    & $1$ \\
         &         & $Sp(2N+2)$ 
                             &           &          &     \\ \cline{2-6}     
         &  $S^{N/2}$ 
                   & $SO(2N+1)$ 
                             & $(1,2,\dots,N-1,{\bf N/2}) $
                                         &.         & $(N-1)^2$ \\ 
         &         & (even $N$)     
                             &           &          &           \\
\cline{2-6} 
         & $S^N$  
                   &  $SO(2N+1)$ 
                             & $(2,4,\dots,2N-2,{\bf N})$ 
                                         &.         & $3N^2-4N+1$\\ 
         &         & (odd $N$)
                             &           &          & \\
\cline{2-6}
         & $\bar F^2 S$
                   &  $SO(2N+3)$
                             & $({\bf 2},2,\dots,2,{\bf 1})$
                                         &.         & $5$ ($N=2$) \\
         &         &         &           &.         & $8$ ($N=3$) \\
         &         &         &           & $U(N-4)$
                                                    & $9$ ($N\ge4$) \\
\cline{2-6}   
         & $\Lambda^{N/2}$  
                   &$SO(2N)$ 
                             & $(1,2,\dots,N-2,N/2-1,{\bf N/2})$ 
                                         &.         & $N^2-4N+1$ \\ 
         &         & (even $N$)
                             &           &          &  $(N>2)$\\
\cline{2-6} 
         & $ \Lambda^N$  
                   & $SO(2N)$ 
                             & $(2,4,\dots,2N-4,N-2,{\bf N}) $
                                         &.        & $3N^2-8N+1$\\
         &         & (odd $N$) 
                             &           &         & \\
\cline{2-6}   
         & $F^{n}F'^{N-n}$
                   & $ SO(2N+2)$ 
                             & $(1,2,\dots,N-1,{\bf n},{\bf N-n}) $
                                         &.         & $(N-1)^2$ 
\\ \cline{2-6}
         & $\bar{F}^2\Lambda$  
                   & $ SO(2N+2)$ 
                             & $({\bf 2},2,\dots,2,1,{\bf 1})$
                                         &.         & $4$ ({$N=3$})\\
         &         &         &           & $U(N-4)$
                                                     & $5$ ($N\ge  4$)\\ 
\cline{2-6}
         & $\bar{F}^2 FF'$ 
                   & $SO(2N+4)$ 
                             & $({\bf 2},2,\dots,2,{\bf 1},{\bf 1})$
                                         &.         & $5$ ($N=2$) \\
         &         &         &           &.         & $8$ ($N=3$) \\
         &         &         &           & $U(N-4)$
                                                    & $9$ ($N\ge 4$) \\
\hline
$SO(2N+1)$
         & $V^2 $  &$SO(2N+3) $
                             & ({\bf 2},2,\dots,2,1)
                                         & $SO(2N-3)$ 
                                                    & 2 ($N\ge 2$)\\ \hline
$Sp(2N)$
         & $V $ 
                   &$Sp(2N+2) $
                             & $({\bf 1},1,\dots,1)$
                                         & $Sp(2N-2)$
                                                    &  1 \\ \hline
$SO(2N)$
         & $V^2$   &$SO(2N+2) $ 
                             & ({\bf 2},2,\dots,2,1,1)
                                         & $SO(2N-4)$ 
                                                    & 2 ($N\ge 3$)\\ \hline
\end{tabular} }
\vskip 1cm
\begin{quote}
{\bf Table 1:} {\small Minimal singlet combinations of massive fundamental 
monopoles when the original gauge group is classical. The symbols for the 
representations of 
the dual group $K_{dual}$ are as follows: $F$ 
is the defining representation of $SU(N)$, while $\Lambda$ and $S$ are the 
antisymmetric and the symmetric products of two $F$'s respectively; $V$ is 
either the defining representation 
of a symplectic group or the vector representation of an orthogonal group; 
finally, a bar on top denotes the complex conjugation. The total magnetic 
charge of the singlet combination is written as a row vector of the
integer coefficients appearing in Eq.~(\ref{gcoeff}), 
ordered according to the Dynkin diagram of
the original gauge group $G$ in Fig.~1, with the $n_j$ indicated by boldface 
type. }
\end{quote}

\vskip 1cm
The other entries in Table~1 with nontrivial $K'$ can all be
determined by studying appropriate embeddings.  As an example, for
$G=SO(k+2)$ broken to $SO(k)$ ($k\ge 4$) there are color-magnetically neutral
solutions containing two massive fundamental monopoles.  To deal with
these, we first consider the case of $SO(6) \rightarrow SO(4)$.  Viewing
this as $SU(4) \rightarrow SU(2)\times SU(2)$, it is not
hard to construct an approximate solution with the two massive
monopoles widely separated that clearly has no invariance group.  The
corresponding moduli space has ${\rm dim}\, SO(4) = 6$ gauge parameters
and $8-6=2$ cloud structure parameters.  By embedding these solutions in
the larger orthogonal groups, we see that for $k\ge 4$, ${\cal N}_{\rm
structure} \ge 2$ and $K' \subseteq SO(k-3)$.  In order that the number of
parameters be consistent with the decompositions of $\bf g$ shown in
the table, these inequalities must be saturated, indicating that the
embeddings give the generic solution.  

Finally, for the combinations where the number of massless monopoles
grows with $N$ the generic solution cannot be obtained by embedding
from a smaller group, and we expect $K'$ to be trivial.

\vskip  1cm

\hskip 8mm
{\small
\begin{tabular}{|l|l|l|l|r|r|} \hline
$K$
        & Singlet  & $G$ 
                             & $e{\bf g} /4\pi$ 
                                         & dim[${\cal M}$] 
                                                    & ${\cal N}_{\rm
structure} 
\ge $ \\ \hline
$SU(2)$ 
        & $ [4]^2$ & $G_2$    & $({\bf 2},3)$ 
                                         & ${\bf 8}+12=20$ 
                                                    & $9$ \\ \cline{2-6}
        & $  F^2$  & $G_2$    & $(1,{\bf 2})$  
                                         & ${\bf 8}+4=12$  
                                                    & $1$ \\ \hline  
$SU(4)$   & $F\bar{F}\Lambda^2$   
                   & $E_6$    & $({\bf 1},2,3,2,{\bf 1},{\bf 2})$ 
                                         & ${\bf 16}+28=44$
                                                    & $13$\\ \hline  
$SU(5)$   & $\bar{F}\bar\Lambda^2$ 
                   & $E_6$    & $({\bf 1},2,3,2,1,{\bf 2}) $
                                         & ${\bf 12}+32=44$ 
                                                    & $8$ \\ \cline{2-6}  
        & $\bar F^3\bar\Lambda$ 
                   & $E_6$    & $({\bf 3},3,3,2,1,{\bf 1})$ 
                                         & ${\bf 16}+36=52$
                                                    & $12$  \\ \cline{2-6}  
        & $ F^2\bar{F}\Lambda$   
                   & $E_7$    & $({\bf 1},2,3,4,3,{\bf 2},{\bf 2})$
                                         & ${\bf 20}+48=68$ 
                                                    & $24$  \\ \hline  
$SU(6)$   & $\Delta^2$    & $E_6$    & $(1,2,3,2,1,{\bf 2}$) 
                                         & ${\bf 8}+36=44$ 
                                                    & $1$ \\ \cline{2-6}  
        & $F^2\Lambda^2$ 
                   & $E_7$    & $(1,2,3,4,3,{\bf 2},{\bf 2})$ 
                                         & ${\bf 16}+52=68$ 
                                                    & $17$ \\ \cline{2-6}
        & $\bar{F}^3 A$ 
                   & $E_7$    & $({\bf 3},3,3,3,2,1,{\bf 1})$ 
                                         & ${\bf 16}+48=64$ 
                                                    & $13$ \\ \cline{2-6} 
        & $F\bar{F}\Lambda^3$ 
                   & $E_8$    & $({\bf 1},2,3,4,5,3,{\bf 1},{\bf 3})$ 
                                         & ${\bf 20}+68=88$ 
                                                    & $33$  \\ \hline 
$SU(7)$   & $\Delta^7$    & $E_7$    & $(3,6,9,12,8,4,{\bf 7}) $ 
                                         & ${\bf 28}+168= 196$ 
                                                    & $120$ \\ \cline{2-6} 
        & $F \Lambda^3 $ 
                   & $E_8$    & $(1,2,3,4,5,3,{\bf 1},{\bf 3}) $
                                         & ${\bf 16}+72=88$ 
                                                    &  $24$ \\ \cline{2-6} 
        & $\Delta^5\bar{F} $
                   & $E_8$    & $({\bf 1},3,5,7,9,6,3,{\bf 5})$
                                         & ${\bf 24}+132=156$ 
                                                    & $84$ \\ \hline 
$SU(8)$   & $\Delta^8$    & $E_8$    & $(3,6,9,12,15,10,5,{\bf 8})$ 
                                         & ${\bf 32}+240=272$ 
                                                    & $177$ \\ \hline
$SO(5)$   &$V\Psi^2$ & $F_4$    & $({\bf 1},2,3,{\bf 2})$ 
                                         & ${\bf 12}+20=32$ 
                                                    & $10$ \\ \hline  
$SO(7)$   & $[14]^2$ & $F_4$    & $({\bf 2},3,4,2)$ 
                                         & ${\bf 8}+36=44$ 
                                                    & $15$ \\ \hline  
$Sp(6)$   & $\Psi^2$ & $F_4$    & $(1,2,3,{\bf 2})$    
                                         & ${\bf 8}+24=32$ 
                                                    & $3$  \\ \hline 
$SO(8)$   & $ [\Psi^{(+)}]^4[\Psi^{(-)}]^{2}$ 
                   & $E_6$    
                              & $({\bf4},5,6,4,{\bf 2},3)$
                                          & ${\bf 24}+72 =96$ 
                                                    & $44$\\ \hline 
$SO(10)$   & $[\Psi^{(+)}]^4$ 
                   & $E_6$    & $({\bf 4},5,6,4,2,3)$ 
                                          & ${\bf 16}+80=96$ 
                                                    & $35$\\ \cline{2-6} 
        & $V[\Psi^{(+)}]^2$ 
                   & $E_7$    & $({\bf 1},2,3,4,3,{\bf 2},2)$
                                          & ${\bf 12}+56=68$ 
                                                    & $11$ \\ \hline  
$SO(12)$   & $[\Psi^{(-)}]^2$ 
                   & $E_7$    & $(1,2,3,4,3,{\bf 2},2)$ 
                                          & ${\bf 8}+60=68$ 
                                                    &  $1$?\\ \cline{2-6}   
        & $V^2[\Psi^{(-)}]^2$ 
                   & $E_8$    & $({\bf 2},3,4,5,6,4,{\bf 2},3)$
                                          & ${\bf 16}+100 =116$ 
                                                    & $34$\\ \hline  
$SO(14)$   & $[\Psi^{(-)}]^4$ 
                   & $E_8$    & $(2,4,6,8,10,7,{\bf 4},5)$  
                                          & ${\bf 16}+168=184$
                                                    & $77$\\ \hline 
$E_6$   & $[27]^3$ & $E_7$    & $({\bf 3},4,5,6,4,2,3)$  
                                          & ${\bf 12}+96=108$  
                                                    & $18$ \\ \hline 
$E_7$   & $[56]^2$ & $E_8$    & $({\bf 2},3,4,5,6,4,2,3)$  
                                          & ${\bf 8}+108=116$ 
                                                    & $1$?\\  \hline
\end{tabular} }

\vskip 1cm
\begin{quote}
{\bf Table 2:} {\small
Minimal singlet combinations of massive fundamental monopoles
when the original gauge group is exceptional. The notation is similar to 
that in Table~1. 
$\Delta$ is the antisymmetric
product of three $F$'s, while $\Psi$ and $\Psi^{(\pm)}$ are spinor
representations of odd- and even-dimensional orthogonal groups. In some
cases, an irreducible representation is denoted by its dimension inside
a square bracket. }
\end{quote}
\vskip 1cm

The results for when the initial gauge group $G$ is exceptional are
summarized in Table 2.  We have used a notation ${\bf 4n}+4q$ for the
dimension of the moduli space, with the boldface numeral indicating
the degrees of freedom associated with the massive monopoles.  Except
for the trivial case $G=G_2$, we do not know the invariant subgroup
$K'$, and so the number in the final column is in general a lower
bound obtained from Eq.~(\ref{shape}) by assuming that $K'$ is
trivial. When this yields a nonpositive number, we have written $1?$
in the last column.

\section{Duality and Threshold Bound States}

The classical BPS multimonopole solutions that we have been studying
can be naturally embedded in an $N=4$ supersymmetric Yang-Mills
theory.  It has been conjectured \cite {dual} that such theories 
possess an exact electric-magnetic duality under which
the spectrum of
electrically charged elementary particles is mirrored by that of the
magnetically charged particles.\footnote{Duality also makes
predictions concerning the dyonic states carrying both electric and
magnetic charges; we do not discuss these here.}  More precisely, the
magnetically charged objects in a theory with gauge group $G$ should
be in one-to-one correspondence with the electrically charged objects
in a theory with the dual group $G_{\rm dual}$.  In the simplest
cases, this correspondence is between the states based on the
elementary quanta and those based on simple soliton solutions.  Thus,
in the $N=4$ supersymmetric theory with $SU(2)$ broken to $U(1)$, the 
states dual to the electrically charged vector mesons are obtained from the 
unit charged monopole and antimonopole solutions.\footnote{If the $SU(2)$ 
theory has only $N=2$ supersymmetry, the unit charged monopoles are actually
dual to quarks in $SU(2)$ doublets. See Ref.~\cite{sethi} for detailed studies
of a conformally invariant model with four families of quarks.}
However, more complex situations
can arise, even when the unbroken group is purely abelian.  Consider,
for example, the case of $SU(3)$ broken to $U(1)^2$.  There are three
electrically charged vector bosons, whose charges in the two unbroken
$U(1)$ factors are (1,0), (0,1), and (1,1); in the BPS limit the mass
of the third of these is the sum of the masses of the first two.  The
duals to the first two objects are the fundamental monopoles of the
theory, but the dual of the third is a threshold bound state of the
two fundamental monopoles.  This state can be constructed
semiclassically by considering the supersymmetric quantum mechanics of
two-monopole systems or, equivalently, by studying a supersymmetric
sigma model on the corresponding moduli space \cite{blum,witten}. In the latter
approach, the bounds states are in one-to-one correspondence with the
harmonic forms on the moduli space that satisfy an appropriate
normalizability condition \cite{sen,porrati}. Such a normalizable harmonic 
form was found recently in Refs.~\cite{klee} and \cite{gaunt}.

Now let us consider the extension of these ideas to theories with
unbroken nonabelian subgroups.  A new feature that arises here is the
presence of massless elementary excitations in the electrically
charged sector.  The duals to these should also be massless, and so
cannot be solitons of the ordinary sort; they are presumably the
massless monopoles that form the nonabelian clouds that we have found.
For the massive particles, on the other hand, the duality picture
should be much closer to that of the MSB case, except that some of the
particles transform under nontrivial representations of the unbroken
gauge group.

As an example, take the case of $SU(N)$ broken to $SU(N-1)\times
U(1)$.  In the electrically charged sector, the $N^2-1$ gauge bosons
of the original group can be decomposed into $N(N-2)$ massless
$SU(N-1)$ gauge bosons, a neutral massless $U(1)$ gauge boson, $(N-1)$
massive bosons with positive $U(1)$ charge belonging to the
fundamental representation of $SU(N-1)$, and $(N-1)$ massive bosons
with negative $U(1)$ charge belonging to the antifundamental
representation of $SU(N-1)$.  As noted above, the duals of the
massless gauge bosons are the massless monopoles and antimonopoles
(except for the case of the neutral bosons, which are self-dual).  The
dual to the positively (negatively) charged massive multiplet is the
fundamental monopole (antimonopole), which, according to the arguments
of the previous section, corresponds to a fundamental
(antifundamental) representation multiplet.

If this group is broken further, to $SU(N-2)\times U(1)^2$, the
elementary particle sector contains two nondegenerate massive
fundamental $SU(N-2)$ multiplets with $U(1)$ charges (1,0) and (0,1);
these are dual to the two kinds of massive fundamental monopoles.
There is also a massive $SU(N-2)$ singlet that carries one unit of
each of the $U(1)$ charges.  Its dual must be a threshold bound state
containing one of each of the fundamental monopoles and $N-1$ massless
monopoles.  Such a state would correspond to a normalizable harmonic
form on the moduli space we discussed in Sec.~4.3; in order to be
unique this form must be either a self-dual or anti-self-dual
$2N$-form \cite{sen}.  With further breaking [e.g., to $SU(N-3)\times U(1)^3$],
additional bound states, containing some monopoles with purely abelian
charges, would also be required.

Other symmetry breaking patterns can be studied in a similar fashion.
In Table~3 we list the breakings of simple groups such that the
unbroken group is a product of a simple group times a product of
$U(1)$ factors and the fundamental monopoles all carry nonabelian
charges.  (The latter requirement implies that every broken simple
root is linked to an unbroken root in the Dynkin diagram.)  For each
of these we have listed the representations of the fundamental
monopoles and indicated the bound states containing such monopoles
that are required by duality.  These examples can in most cases be
embedded in larger groups, in which case there will also be purely
abelian fundamental monopoles and additional bound states containing
these.

In principle, all the bound states listed in Table~3 must be realized
as harmonic forms on appropriate moduli spaces.  However, actually
finding these forms is a rather nontrivial problem.  For the case of
distinct fundamental monopoles with the moduli space metric given
in Ref.~\cite{klee2}, Gibbons \cite{middle} recently gave an answer for
the threshold bound state in the MSB case: a middle form constructed
as a wedge product of a number of harmonic two-forms that are
associated with the Killing vectors $\partial/\partial\psi_A$. One
might have hoped that this construction would carry over to the
present NUS limit and produce the expected harmonic $2N$-form on the
moduli space for two massive and $N-1$ massless monopoles of
$Sp(2N+2)$ or $SU(N+2)$ broken to $SU(N) \times U(1)^2$.
Unfortunately, this is not the case. Although the harmonicity of the
middle form is likely to be preserved, the normalizability is
not. Further, this middle-form is invariant only under the Cartan
subgroup of the unbroken gauge group $SU(N)$, implying that the
corresponding state is electrically charged and cannot be the purely
magnetic threshold bound state.  This difficulty is compounded by the
fact that even the MSB moduli space metric is unknown for most cases.

Finally, we want to emphasize the fact that some of the
required bound states transform nontrivially under the unbroken gauge
group.  As we have noted, there are pathologies associated with
configurations that have nonzero long-range nonabelian magnetic
fields, and a meaningful moduli space emerges only if we insist that
the total magnetic charge be purely abelian.  Since the non\-singlet
bound states necessarily involve only some of the monopoles described
by the moduli space, the corresponding harmonic forms cannot be
normalizable in the usual sense.

\newpage

\vskip 5mm
\hskip -3mm
{\small
\begin{tabular}{|l|l|l|l|r|}\hline
$G$ &$G_{dual}$ & Unbroken Dual Group  & Massive
Monopoles     & Bound States \\ \hline
$SU(N+1)$& 
$SU(N+1)$  & $SU(N)\times U(1)$     & $F$  & None \\ \cline{3-5}
&           & $U(1)\times SU(N-1)\times U(1)$     & $\bar{F}$, $F$ & 
           $F\bar{F}\Rightarrow [1]$ \\ \hline
$SO(2N+1)$&
$Sp(2N)$   & $U(1)\times Sp(2N-2)$    & $V$        & $VV\Rightarrow [1]$
           \\ \cline{3-5}
&           & $SU(N)\times U(1)$     & $S$      & None \\ \cline{3-5}
&           & $U(1)\times SU(N-1)\times U(1)$
                                    & $\bar{F}$, $S$
                                                 & $\bar{F}S\Rightarrow F$ \\
&           & & & $\bar{F}\bar{F} S\Rightarrow [1]$ \\ \hline
$Sp(2N)$&
$SO(2N+1)$ & $U(1)\times SO(2N-1)$  & $V$        & None \\ \cline{3-5}
&           & $SU(N)\times U(1)$     & $F$        & $FF\Rightarrow \Lambda$ 
           \\ \cline{3-5}
&           & $U(1)\times SU(N-1)\times  U(1)$ 
                                    & $F$, $\bar{F}$   
                                                 & $F\bar{F}\Rightarrow [1]$\\
&           & & &  $FF \Rightarrow \Lambda$ \\
&           & & &  $FF\bar{F}  \Rightarrow F$ \\ \hline
$SO(2N)$&
$SO(2N)$   & $U(1)\times SO(2N-2)$  & $V$        & None \\ \cline{3-5}
&           & $SU(N)\times U(1)$     & $\Lambda$ & None \\ \cline{3-5}
&           & $SU(N-1)\times U(1)^2$ & $F$, $F'$  & $FF'\Rightarrow \Lambda$ 
           \\ \cline{3-5}
&           & $U(1)\times SU(N-1)\times U(1)$
                                    & $\bar F$, $\Lambda$
                                                 & $\bar F\Lambda\Rightarrow F$
           \\ \cline{3-5}
&           & $U(1)\times SU(N-2)\times U(1)^2$
                                    & $F$, $F'$, $\bar{F}$ 
               & $F\bar F \Rightarrow [1]$\\
&           & & & $F'\bar F \Rightarrow [1]$\\                                 
&           & & & $FF'\Rightarrow \Lambda$ \\
&           & & &  $FF'\bar{F}\Rightarrow F$\\ \hline
$G_2$  &
$G_2$  & $U(1)\times SU(2)$   & $[4]$        & $[4]\times [4]\Rightarrow [1]$ 
       \\ \cline{3-5}(${\rm  dim}=14$) 
&       & $SU(2)\times U(1)$   & $[2]$        & $[2]\times [2]\Rightarrow [1]$
       \\
&       & & & $[2]\times [2]\times [2]\Rightarrow [2]$\\ \hline
$F_4$  &
$F_4$  & $Sp(6)\times U(1)$   & $[14]$       & $[14]\times[14] \Rightarrow [1]$
       \\ \cline{3-5} (${\rm dim}=52$) 
&       & $U(1)\times SO(7)$   & $\Psi$       & $\Psi\Psi\Rightarrow V$
       \\  \cline{3-5}
&       & $U(1)\times SO(5)\times  U(1)$
                              & $V$, $\Psi$  & $\Psi\Psi\Rightarrow [1]$\\
&       & & & $V\Psi\Rightarrow \Psi$\\
&       & & & $V\Psi\Psi\Rightarrow V$\\
&       & & & $VV\Psi\Psi\Rightarrow [1]$\\ \hline
$E_6$  &
$E_6$  & $U(1)\times SU(6)$  & $\Delta$   & $\Delta\Delta
                                                \Rightarrow [1] $
       \\ \cline{3-5} (${\rm dim}=78$)
&       & $SO(10)\times U(1)$  & $\Psi^{(+)}$ & None\\ \cline{3-5}
&       & $U(1)\times SO(8)\times U(1)$
                              & $V$, $\Psi^{(-)}$
                                             & $V\Psi^{(-)}\Rightarrow   
         \Psi^{(+)}$ \\ \cline{3-5}
&       & $U(1)\times SU(4)\times U(1)^2$
         & $\bar F$, $F$, $\Lambda$
           & $\bar F F\Rightarrow [1]$\\
&       & & & $\bar F\Lambda\Rightarrow F$\\
&       & & & $F \Lambda\Rightarrow\bar F$\\
&       & & & $\bar F F\Lambda\Rightarrow \Lambda$\\
&       & & & $\bar F F\Lambda\Lambda\Rightarrow [1]$\\ \hline
\end{tabular}  }

\vskip 1cm
\hskip 5mm {\small
\begin{tabular}{|l|l|l|l|r|}\hline
$G$ & $G_{dual}$  & Unbroken Dual Group   & Massive
Monopoles     & Bound States \\ \hline
${E}_7$  &
${E}_7$  & $U(1)\times  {E}_6$    & $[27]$     & None \\ \cline{3-5}
(${\rm dim}=133$)
&       & $U(1)\times SU(7)$   & $\Delta$   & $ \Delta\Delta
                                                \Rightarrow \bar{F}$
       \\ \cline{3-5} 
&       & $SO(12)\times U(1)$  & $\Psi^{(-)}$ & $\Psi^{(-)}\Psi^{(-)}
                                                \Rightarrow [1] $ 
       \\ \cline{3-5}
&       & $U(1)\times SO(10)\times U(1)$
                              & $V$, $\Psi^{(+)}$
                                             & $V\Psi^{(+)}\Rightarrow   
         \Psi^{(-)}$ \\
&       & & & $V\Psi^{(+)}\Psi^{(+)}\Rightarrow [1]$\\  \cline{3-5}
&       & $U(1)\times SU(6)\times U(1)$
         & $\bar F$, $\Delta$ 
           & $\bar F \Delta\Rightarrow \Lambda $ \\
&       & & & $\Delta\Delta \Rightarrow [1] $ \\
&       & & & $\bar F \Delta\Delta\Rightarrow \bar F $\\ \cline{3-5}
&       &$ U(1)\times SU(5)\times U(1)^2$
         & $\bar F$, $F$, $\Lambda$
           & $\bar F F\Rightarrow [1]$\\
&       & & & $\bar F\Lambda\Rightarrow F$\\
&       & & & $F\Lambda\Rightarrow \bar\Lambda$\\
&       & & & $F\Lambda\Lambda\Rightarrow [1]$\\
&       & & & $\bar F F \Lambda\Rightarrow \Lambda $ \\
&       & & & $\bar F F \Lambda\Lambda\Rightarrow \bar F$\\
&       & & & $\bar F F F\Lambda\Lambda\Rightarrow [1] $\\ \hline     
$ E_8$  & 
$ E_8$  &  $U(1)\times {E}_7$    & $[56]$   & $[56]\times[56] \Rightarrow [1]$
          \\ \cline{3-5}(${\rm dim}=248$) 
&       &  $U(1)\times SU(8)$  & $\Delta$ & $\Delta\Delta\Rightarrow
          \bar{\Lambda}$ \\ 
&       & & & $\Delta\Delta\Delta\Rightarrow F$\\\cline{3-5}
&       &  $SO(14)\times U(1)$  & $\Psi^{(-)}$ & $\Psi^{(-)}\Psi^{(-)}
          \Rightarrow V$ \\ \cline{3-5}
&       &  $U(1)\times SO(12)\times U(1)$
         & $\Psi^{(-)}$, $V$  
           &  $\Psi^{(-)}\Psi^{(-)}\Rightarrow [1]$\\
&       & & &  $\Psi^{(-)}V\Rightarrow \Psi^{(+)}$ \\
&       & & &  $\Psi^{(-)}\Psi^{(-)}V\Rightarrow V$\\
&       & & &  $\Psi^{(-)}\Psi^{(-)}VV\Rightarrow [1]$ \\ \cline{3-5}
&       & $U(1)\times SU(7)\times U(1)$
         & $\bar F$, $\Delta$
           & $\bar F\Delta\Rightarrow \Lambda $\\
&       & & & $\Delta\Delta\Rightarrow \bar F$ \\
&       & & & $\bar F \Delta\Delta\Rightarrow \bar\Lambda$ \\
&       & & & $\bar F \Delta\Delta\Delta\Rightarrow F$\\
&       & & & $\bar F\bar F\Delta\Delta\Delta\Rightarrow [1]$\\  \cline{3-5}
&       & $U(1)\times SU(6)\times U(1)^2$
         & $\bar F$, $F$, $\Lambda$
           & $\bar F F\Rightarrow [1]$\\
&       & & & $\bar F\Lambda\Rightarrow F$\\
&       & & & $F\Lambda\Rightarrow\Delta$\\
&       & & & $F\Lambda\Lambda\Rightarrow\bar F$\\
&       & & & $\bar F F\Lambda\Rightarrow\Lambda$\\
&       & & & $\bar F F\Lambda\Lambda\Rightarrow\bar \Lambda$\\
&       & & & $F F\Lambda\Lambda\Rightarrow [1]$\\
&       & & & $\bar F F \Lambda\Lambda\Lambda\Rightarrow [1]$\\ 
&       & & & $\bar F F F\Lambda\Lambda\Rightarrow \bar F$\\
&       & & & $\bar F F F\Lambda\Lambda\Lambda\Rightarrow \bar F$\\
&       & & & $\bar F \bar F F F\Lambda\Lambda\Rightarrow [1]$\\
\hline
\end{tabular} }
\vskip 1cm

\begin{quote}
{\bf Table 3:} {\small Representations of massive fundamental monopoles and
their threshold bound states. The notation for representations follows
that of Tables 1 and 2.}
\end{quote}

\section{Conclusion}

In this paper we have used the multimonopole moduli space as a tool
for investigating the properties of monopoles carrying nonabelian
magnetic charges.  If the net magnetic charge is purely abelian, the
moduli space for the case with an unbroken nonabelian subgroup can be
obtained as a smooth limit of that for the MSB case.  In this limit
the moduli space describes multimonopole solutions that are composed
of  one or
more color-magnetically neutral combinations of monopoles.  In each of
the latter there are are a number of massive fundamental monopoles,
corresponding to embeddings of the $SU(2)$ monopole, that carry both
abelian and nonabelian magnetic charge.  These are surrounded by a
cloud within which there is a nonzero nonabelian magnetic field.

By studying the approach to the NUS limit, we are led to interpret
this cloud as being composed of massless monopoles carrying purely
nonabelian magnetic charges.  These can be understood as limits of the
fundamental monopoles of the MSB case that correspond to simple roots
of the unbroken nonabelian subgroup.  However, they differ from the
other fundamental monopoles in that there is no classical soliton
corresponding to an isolated massless monopole.  When they
coalesce to form a nonabelian cloud, they lose their identity as
individual objects.  Thus, although the number of parameters remains
unchanged as one goes from the MSB case to the NUS case, the position
and $U(1)$ orientations of these monopoles are transformed into gauge
orientation and structure parameters describing the cloud as a whole.

There are a number of outstanding issues to be addressed.  We have
worked entirely within the context of the BPS limit. To what extent do
our results apply to models (such as realistic grand unified theories)
that have nonvanishing Higgs potentials?  Such models will still have
a number of massive fundamental monopoles belonging to representations
of the dual of the unbroken gauge group.  At least for Higgs masses small
compared to the vector meson masses, the leading effects of the
departure from the BPS limit could be incorporated by adding to the
moduli space Lagrangian a potential energy depending on the monopole
separations and the cloud structure parameters.   Presumably at least
some of the color-magnetically neutral combinations of monopoles
are stably bound (both classically and quantum mechanically) by this
potential energy \cite{vachaspati}, since the Brandt-Neri-Coleman analysis
\cite{brandt} shows that stable 
configurations with large nonabelian magnetic charge are impossible.

Another important question is that of how our largely classical
analysis must be modified to take into account quantum effects.  We
discussed briefly in Sec.~3 the quantization of the moduli space
coordinates and the nature of the low energy eigenstates of the moduli
space Hamiltonian.  However, we have not addressed at all the question
of how the moduli space itself might be modified by quantum
corrections.  (Note that the BPS limit can be maintained under quantum
corrections in theories with extended supersymmetry.)  
For example, at the classical level the energy does not depend on the
values of the cloud structure parameters, but the corresponding degeneracy
does not seem to be
required by the BPS conditions at the quantum level.  Does this mean
that one-loop
effects modify the low energy moduli space Lagrangian?  It would be
clearly desirable to go beyond the semiclassical approximation and
make a connection with the work of Seiberg and Witten \cite{seiberg}.
One would also like to understand what the
effects of confinement on nonabelian magnetic charges 
are and how they should be incorporated.

Perhaps most interesting are the questions connected with the duality
hypothesis.  Particularly intriguing is the role of the massless
monopoles, which are naturally recognized as being the objects that
are dual to the massless gauge bosons carrying electric-type color
charges.  In fact, 
if the electric and magnetic sectors are to be on an equal setting,
the full multiplet of gauge bosons should have a counterpart
comprising not only the massless monopoles and antimonopoles, but also
neutral gauge particles corresponding to the Cartan subalgebra.  In
one sense, the latter should be seen as being their own dual, just as
the photon is in the $SU(2)\rightarrow U(1)$ case.  However, the fact
that the choice of the Cartan subalgebra for the unbroken group is not
gauge-invariant shows that the particular separation into monopoles,
antimonopoles, and self-dual objects is to some extent arbitrary.
Clearly, there is much to be learned about these objects.  Indeed, one
might hope that a fuller understanding of these massless monopoles
could form the basis for a dual approach to nonabelian interactions
that would prove complementary to that based on the perturbative gauge
bosons.

\vskip 1cm
\centerline{\large\bf Acknowledgments}
\vskip  5mm

We thank Gordon Chalmers, Calin Lazaroiu, and Jaemo Park for helpful
conversations, and Michael Murray for communicating his results to us.
This work was supported in part by the U.S. Department of
Energy. K.L. is supported in part by the NSF Presidential Young
Investigator program.

\vskip 1cm
\setcounter{equation}{0}
\makeatletter
\renewcommand\theequation{A.\arabic{equation}}
\makeatother

\leftline{\Large\bf Appendix: Triholomorphic $SU(N)$ Isometry}
\vskip 5mm
\noindent
We start with the observation that the $E_{AB}$'s preserve $\sum_A {\bf r}_A$,
which is the relative position vector between the two massive monopoles.
This can be seen by rewriting the vector field in  three-dimensional
coordinates:
\begin{equation}
E_{AB}=e^{-i(\psi_A-\psi_B)/2}\left[ f^{(a)}_{AB}\left(
\frac{\partial}{\partial r_A^{a}}-\frac{\partial}{\partial r_B^{a}}\right)
+g_{AB}\frac{\partial}{\partial\psi_A}+g_{BA}^*\frac{\partial}{\partial\psi_B}
\right]. \label{newform}
\end{equation}
(The details of the $N\times N$ matrices $f^{(a)}$ and $g$ will not matter
here.)  
Recalling that the scalar quantity $\sum_A r_A$ is also invariant,
one can easily see that the metric in Eq.~(\ref{metric}) is invariant if and 
only if the Lie derivative of the one-form
\begin{equation}
\Omega\equiv \sum_A r_A\,(d\psi_A+\cos\theta_A\,d\psi_A)=
\frac{i}{4}\sum_A \left(\xi_A^*d\xi_A-\xi_Ad\xi_A^*-\zeta_A^*d\zeta_A+
\zeta_Ad\zeta_A^*\right), \label{V}
\end{equation}
vanishes. The Lie derivative of the differential form can be succinctly written
as
\begin{equation}
{\cal L}_{E_{AB}}\Omega=d\langle E_{AB},\Omega \rangle+\langle E_{AB},d
\Omega\rangle.
\end{equation}
The two terms cancel each other with $\Omega$ given as in Eq.~(\ref{V}) 
(this is easiest to see in complex coordinates),
so the $E_{AB}$'s are indeed Killing vector fields.

To show that $E_{AB}$ is triholomorphic, we compute the Lie derivative of
the k\"ahler form $w^{(a)}_{{\cal M}_0}$,
\begin{equation}
{\cal L}_{E_{AB}}w^{(a)}_{\cal M}=d\langle E_{AB},w^{(a)}_{\cal M}
\rangle+\langle E_{AB},dw^{(a)}_{\cal M}\rangle.
\end{equation}
The k\"ahler forms are closed, so that the second term is null, while the 
first term is 
\begin{equation}
d\langle E_{AB},w^{(a)}_{{\cal M}_0}\rangle=\frac{g^2\lambda}{8\pi}
d\langle E_{AB},w^{(a)}_{\rm flat}\rangle-\frac{\bar \mu}{2}
d\left\langle E_{AB},\epsilon^{abc}\left(\sum_A dr^b_A\right)
\wedge \left(\sum_B dr_B^c\right)\right\rangle.
\end{equation}
Because $E_{AB}$ is orthogonal to $\sum_A dr^a_A$, the $\bar \mu$ dependent
term vanishes identically. Then,
\begin{equation}
{\cal L}_{E_{AB}}w^{(a)}_{{\cal M}_0}=\frac{g^2\lambda}{8\pi}
d\langle E_{AB},w^{(a)}_{\rm flat}\rangle=
\frac{g^2\lambda}{8\pi}{\cal L}_{E_{AB}}w^{(a)}_{\rm flat}=0.
\end{equation}
This concludes the proof that the $E_{AB}$'s are triholomorphic
Killing vector fields. It follows that the NUS metric in Eq.~(\ref{metric}) 
admits a $U(N)$ isometry that preserves its hyperk\"ahler structure.

\end{document}